\documentclass[11pt,a4paper]{article}
\pdfoutput=1
\usepackage{jcappub}
\usepackage[british]{babel}
\usepackage{latexsym,amsmath,amssymb}
\usepackage{graphicx}
\usepackage{booktabs}
\usepackage{multirow}
\usepackage{lpic}

    \newcommand{\be}{\begin{eqnarray}}
    \newcommand{\ee}{\end{eqnarray}}
    \newcommand{\bea}{\begin{eqnarray}}
    \newcommand{\eea}{\end{eqnarray}}
    \newcommand{\ba}{\begin{array}}
    \newcommand{\ea}{\end{array}}

    \renewcommand{\H}{{\cal H}}
    \newcommand{\ceff}{c_{\rm eff}}

    \renewcommand{\(}{\left(}
    \renewcommand{\)}{\right)}
    \renewcommand{\[}{\left[}
    \renewcommand{\]}{\right]}

    \newcommand{\lcb}{\left\{}
    \newcommand{\rcb}{\right\}}

    \renewcommand{\aa}{e_\pi}
    \newcommand{\ff}{f_\pi}
    \let\mug\gg
    \renewcommand{\gg}{g_\pi}

\newcommand{\Mpc}{\text{Mpc}}

\title{The traces of anisotropic dark energy in light of Planck\footnote{Based on observations obtained with Planck (\href{http://www.esa.int/Planck}{http://www.esa.int/Planck}), an ESA science mission with instruments and contributions directly funded by ESA Member States, NASA, and Canada.}}

\author[a]{Wilmar Cardona,}
\author[b,c,d]{Lukas Hollenstein,}
\author[a,e]{Martin Kunz}

\affiliation[a]{D\'epartement de Physique Th\'eorique and Center for Astroparticle Physics, Universit\'e de Gen\`eve, 24 Quai Ernest Ansermet, 1211 Gen\`eve 4, Switzerland}
\affiliation[b]{IAS Institute of Applied Simulation, ZHAW Zurich University of Applied Sciences, Gr\"uental, PO\ Box, 8820 W\"adenswil, Switzerland}
\affiliation[c]{Institut de Physique Th\'eorique, CEA-Saclay, Orme des Merisiers bat.\ 774, PC 136, 91191 Gif-sur-Yvette Cedex, France}
\affiliation[d]{Institute of Theoretical Astrophysics, University of Oslo, 0315 Oslo, Norway}
\affiliation[e]{African Institute for Mathematical Sciences, 6 Melrose Road, Muizenberg, 7945, South Africa}

\emailAdd{wilmar.cardona@unige.ch}
\emailAdd{lukas.hollenstein@zhaw.ch}
\emailAdd{martin.kunz@unige.ch}

\abstract{We study a dark energy model with non-zero anisotropic stress, either linked to the dark energy density or to the dark matter
density. We compute approximate solutions that allow to characterise the behaviour of the dark energy model and to assess the stability
of the perturbations. We also determine the current limits on such an anisotropic stress from the cosmic microwave background data by the Planck
satellite, and derive the corresponding constraints on the modified growth parameters like the growth index, the effective Newton's constant and the
gravitational slip.}

\begin{document}
\maketitle
\flushbottom

\section{Introduction}

The last twenty years have witnessed a revolution in observational cosmology, with 
an incredible growth of data available to cosmologists. When interpreted within 
the cosmological standard model, one consequence of the observations is the need 
for an accelerated expansion of the Universe. To drive this acceleration a new 
constituent is required, called dark energy. The main candidate model for the dark 
energy is the cosmological constant $\Lambda$, but this model suffers from severe 
fine-tuning issues. Even though cosmologists have been very active and have 
invented a large number of other possible models, including modifications to 
general relativity as the theory of gravity, none of them appear like natural 
candidates for the dark energy (see e.g.\ \cite{Copeland:2006wr, Durrer:2008in, 
Frieman:2008sn, Amendola2010, Clifton:2011jh, Amendola:2012ys} for reviews).

The jury is therefore still out concerning the nature of the dark energy, and it 
may be preferable to approach the problem from the observational side, by 
characterising the possible observational consequences of the dark energy, and 
then investigating the link between those and its physical nature. (See e.g.\ 
\cite{Kunz:2012aw} for a short review, as well as \cite{Battye:2012eu, 
Sawicki:2012re, Baker:2012zs, Gubitosi:2012hu, Bloomfield:2012ff} for recent 
works on parameterised or effective action approaches.)

Useful quantities that are close to the observations are the functions that 
describe the metric \cite{Amendola:2007rr, Hu:2007pj, Sawicki:2012re, 
Amendola:2012ky, Motta:2013cwa}. If we only use quantities up to first order in
perturbation theory, and keep only scalar perturbations, then the metric can be 
written as
\be
g_{\mu\nu}dx^\mu dx^\nu = a^2\left\{ -\(1+2\psi\)\,d\eta^2
  + \(1-2\phi\)\,\delta_{ij} dx^i dx^j \right\}\,, \label{eq:metric}
\ee
where we used the longitudinal gauge. The relevant quantities then are the scale 
factor $a(\eta)$, or equivalently the Hubble parameter $H(\eta)$, and the two 
gravitational potentials $\phi(k,\eta)$ and $\psi(k,\eta)$. The evolution of the 
Hubble parameter is measured by probes like the luminosity distance to type-Ia 
supernovae (SN-Ia) or the baryonic acoustic oscillations (BAO). Possible probes of 
the gravitational potentials include weak lensing which measures the integral of 
$\phi+\psi$, the motion of test particles which is governed by $\psi$ or also the 
integrated Sachs-Wolfe (ISW) effect of the cosmic microwave background (CMB) or 
the large-scale distribution of galaxies.

The standard dynamical dark energy model invokes an additional minimally coupled scalar 
field, possibly with a non-canonical kinetic term. An important feature of this 
class of models is that the scalar field does not support any anisotropic stress 
in linear theory, i.e.\ the space-space part of its energy-momentum tensor has only a 
trace contribution. So-called modified-gravity models, which include 
scalar-tensor, $f(R)$, brane-world and similar models, generically have a non-zero 
(effective) contribution to the anisotropic stress. As a non-zero anisotropic 
stress manifests itself through a gravitational slip, $\phi\neq\psi$, the 
effective anisotropic stress provides a crucial observational test for the nature 
of the dark energy \cite{Kunz:2006ca, Saltas:2010tt}.

Much of the effort in the literature has so far focused on determining 
observational bounds on the background evolution, usually for scalar field models 
without anisotropic stress (e.g.\ \cite{Copeland:2006wr, Frieman:2008sn,Sapone:2010iz,Amendola:2012ys}). 
In this paper we will investigate specifically how a non-zero 
anisotropic stress impacts the dark energy and dark matter perturbations, as well 
as the CMB. For this, we use phenomenological prescriptions that are 
motivated by the typical behaviour of the anisotropic stress for a range of 
modified gravity models. We focus on two model ingredients: externally and 
internally sourced anisotropic stress which reflects a simplified version of a 
more general structure proposed in \cite{Sawicki:2012re}. The paper is structured 
as follows: in the next section we briefly present the perturbation equations 
including anisotropic stress, which also serves to define our notation, as well as 
our closure relations for the pressure perturbations and the dark energy 
anisotropic stress. We then study the phenomenological impact of the presence of a 
nonzero anisotropic stress in section 3, before discussing observational 
constraints from the CMB and geometrical probes in section 4. In section 5 we 
relate the effect of the anisotropic stress to the `modified growth' 
parameterisations that are commonly used in the literature.  We finally conclude 
in section 6. The appendices contain more detailed explanations for the stability 
analysis as well as some exact but cumbersome solutions of the perturbation 
evolution.


\section{Models of anisotropic dark energy}

\subsection{Perturbation equations}

We have already given the perturbed metric in longitudinal gauge in Eq.\ 
(\ref{eq:metric}). A prime will stand for the derivative w.r.t.\ conformal time, 
$\eta$, and $\H\equiv a'/a=aH$ is the comoving Hubble parameter while $H$ is the 
physical Hubble parameter that takes the value of the Hubble constant $H_0$ today 
when $a_0=1$. The continuity and Euler equations for the dark energy 
perturbations read \cite{Bardeen:1980kt, Kodama:1985bj, Ma:1995ey}
\be
\delta'_{de} +3\H\(\frac{\delta P_{de}}{\rho_{de}}-w\delta_{de}\)
  +(1+w)kv_{de}  -3(1+w)\phi'  =  0
\label{eq:de-cont}
\\
v_{de}' +\H(1-3c_a^2)v_{de} -k\( \psi +\frac{\delta P_{de}}{(1+w)\rho_{de}}
-\frac{2\pi_{de}}{3(1+w)} \)  =  0
\label{eq:de-eul}
\ee
where the adiabatic sound speed is
\be
  c_a^2  \equiv  \frac{P_{de}'}{\rho_{de}'}  =  w - \frac{w'}{3\H(1+w)} \,.
\ee
The evolution equations for the dark matter are the same, but with 
$w_m=\delta P_m = \pi_m = 0$. Notice that in terms of the often used variable 
$\sigma$ for the anisotropic stress \cite{Ma:1995ey} we have that 
$\pi = (3/2)(1+w)\,\sigma$.

In addition to these evolution equations we need the Einstein constraint 
equations to compute the impact of the dark matter and dark energy perturbations 
on the metric. For the scalar perturbations considered here, there are two 
independent Einstein equations which we can take to be
\be
- k^2 \phi &=& 4 \pi G a^2 \left(\rho_m \Delta_m + \rho_{de} \Delta_{de} \right) 
\, , \label{eq:poisson} \\
k^2 (\phi - \psi) &=& 8 \pi G a^2 \rho_{de} \pi_{de} \, . \label{eq:aniso}
\ee
Here we wrote the Poisson equation (\ref{eq:poisson}) directly in terms of the 
comoving density perturbation $\Delta$ which is linked to the density 
perturbation in the longitudinal gauge $\delta$ by a gauge transformation, 
$\Delta = \delta + 3 \H (1+w) v/k$. In the equation for the slip (\ref{eq:aniso}) 
we further used that $\pi_m = 0$ (which is strictly speaking only true at first 
order in perturbation theory \cite{Ballesteros:2011cm}). From the two equations 
(\ref{eq:de-cont}) and (\ref{eq:de-eul}) one can derive a single second order 
evolution equation for $\delta_{de}$ by solving the continuity equation 
(\ref{eq:de-cont}) for $v_{de}$ and substituting that (and its derivative) into 
the Euler equation (\ref{eq:de-eul}). We find
\be
  \delta''_{de} +\(1-6w\) \H \delta'_{de}
    +3 \H \(\!\frac{\delta P_{de}}{\rho_{de}}\!\)'
    +3 \Big[(1-3w)\H^2+\H'\Big] \(\frac{\delta P_{de}}{\rho_{de}}-w\delta_{de}\)
    -3 \H w' \delta_{de}
\nonumber \\
 = 3(1+w)\[\phi'' + \(1-3w +\frac{w'}{(1+w)\H}\) \H \phi' \]
    - k^2\[ (1+w)\psi + \frac{\delta P_{de}}{\rho_{de}} -\frac{2}{3}\pi_{de} \] 
\,.
\label{eq:de-secondorder}
\ee
To this point we did not make any assumptions on $\delta P_{de}$, $\pi_{de}$ and 
$w$. However, already the last term makes clear that $k^2\pi_{de}$ acts as a 
source for $\delta_{de}$ while the pressure counteracts the gravitational 
collapse. $-k^2\psi$ is also a source because $\psi=\phi$ for vanishing 
anisotropic stresses and $-k^2\phi\propto \H^2\Delta_{tot}$.

\subsection{Modelling the DE pressure perturbation}

We define the effective, non-adiabatic sound speed of DE in its rest-frame, 
$\partial_\mu P_{de}\equiv c_s^2 \partial_\mu \rho_{de}$. This is the form of 
the sound speed that e.g.\ $K$-essence type models exhibit, with $c_s^2=1$ for a 
canonical scalar field. When we perform a gauge transformation to the 
longitudinal gauge, we find
\be
\frac{\delta P_{de}}{\rho_{de}} = c_s^2\delta_{de} +3(1+w)\(c_s^2 - c_a^2\) 
    k^{-1}\H v_{de} \,. \label{eq:dp}
\ee
We keep the sound speed $c_s$ as a free parameter, but assume it to be a 
constant.

\subsection{Model 1: externally sourced anisotropic stress}

In the quasi-static limit of DGP, the metric potentials are directly linked to 
the matter  perturbations through a time-dependent function \cite{Koyama:2005kd} 
and consequently also the anisotropic stress is proportional to $\Delta_m$ 
\cite{Kunz:2006ca} with, in general, a time-dependent coefficient. Another 
motivation to link the dark energy anisotropic stress to the matter is the 
possibility of couplings between dark energy and dark matter. To keep the model 
simple we use
\be
  \pi_{de}\ \equiv\  \aa a^n \Delta_m
\label{eq:model:1}
\ee
with a constant coefficient $\aa$. We will see that this term will act as an 
additional source for $\delta_{de}$. When looking at the constraints from 
data in section \ref{section:4} we will fix $n=0$, which is also roughly the behaviour of the effective
anisotropic stress in the DGP model.

\subsection{Model 2: counteracting the pressure perturbation}

In \cite{Song:2010rm} a coupling of the anisotropic stress to the pressure 
perturbation was proposed, $\sigma \propto \delta P_{de}/\rho_{de}$, linking 
isotropic and anisotropic stresses which appears quite natural\footnote{A similar 
link was also exploited in \cite{Kunz:2006ca} to define the pressure perturbation when mimicking DGP.}. For non-zero sound 
speed the pressure perturbation is related to the density perturbation by our 
model (\ref{eq:dp}). Here we formulate the dependence directly in terms of the 
comoving density perturbation since this is a gauge-invariant prescription. In 
addition, we allow for a different behaviour on small and large scales, with a 
transition scale $k_T$
\be
\pi_{de}\ = \ff \frac{(k/k_T)^2}{1+(k/k_T)^2} \Delta_{de} \quad , \, k_T = \gg
\H(a)
\ee
with constant parameters $\ff$ and $\gg$. We can then write this model also as
\be
\pi_{de}\ = \frac{\ff}{1 +(\gg\H/k)^2} \Delta_{de} \, .
\label{eq:model2}
\ee
For $\ff = (3/2) c_s^2$ the anisotropic stress cancels the pressure perturbation 
in the Euler equation (\ref{eq:de-eul}) on sub-horizon scales, but not in the 
continuity equation (\ref{eq:de-cont}) and the Einstein constraints.

The dark energy model used here corresponds actually to a subset of the closure 
relations given in Eq.\ (2.47) of  \cite{Sawicki:2012re}, although we originally 
started this work before those relations were derived. The pressure perturbation 
is just the first term of the first equation in their (2.47) with $c_s^2=C^2$ 
(plus the usual contribution to $\Sigma_1$ from the gauge transformation). The 
externally sourced anisotropic stress contribution parameterised by $\aa$ belongs 
in this context to the dark matter coupling term with parameter $\beta_\pi$. The 
second contribution to the anisotropic stress here corresponds to the first term 
in their (2.47), parameterised by $\Pi$. The scale-dependence in our prescription 
leads to a suppression on large scales, and then `turns on' the anisotropic 
stress on scales $k \mug k_T$, similar to the behaviour of the non-minimally 
coupled K-\emph{essence} model described in the second part of \cite{Sawicki:2012re} 
where the authors found a `perfect' and an `imperfect' regime. (However, 
here we limit ourselves to a case where effectively $M_C^2 = 0$ and $\kappa'=0$.) 
For a detailed comparison to \cite{Sawicki:2012re}, notice that we use a 
different sign convention for the metric, and that their $(k/a)^2 \delta\pi$ is 
our $\rho_{de} \pi_{de}$. 

Our model also satisfies the constraint equations derived in \cite{Baker:2012zs}. These are a consequence
of the Bianchi identities, which lead to $\nabla_\mu G^\mu_\nu=0$, and of the covariant 
conservation of the matter energy-momentum tensor $\nabla_\mu T^\mu_\nu=0$. The constraint
equations are equivalent to the covariant conservation of the energy momentum tensor of the
dark energy. For a general fluid they are equivalent to the conservation equations (\ref{eq:de-cont}) and (\ref{eq:de-eul}).

Anisotropic stress perturbations in dark energy have been studied
before, see e.g.\ Refs.\ \cite{Koivisto:2005mm, Mota:2007sz,
Sapone:2012nh, Sapone:2013wda}. However, note that the approach taken
in these references is very different to ours (and that of Refs.\
\cite{Kunz:2006ca, Song:2010rm}). In the former, the Boltzmann hierarchy of a
generic fluid of collisional particles is truncated at the level of
the anisotropic stress \cite{Hu:1998kj}. A viscosity parameter
$c_\text{vis}^2$ is introduced and the behaviour of anisotropic stress
of radiation (up to the quadruple) is recovered for
$c_\text{vis}^2=1/3$. It turns out that such anisotropic stress, often
referred to as viscosity, tends to wash out fluctuations in the dark
energy and, therefore, makes dark energy perturbations even harder to
detect than in the absence of anisotropic stress. On the contrary, the
models discussed here are designed to imitate typical modified gravity
scenarios and therefore aim at creating very different effects, e.g.\
detectable gravitational slip on sub-horizon scales.



\section{Phenomenology}\label{section:3}

From now on we consider  the equation of state $w$ as a free parameter, but assume it to be a constant. From the evolution equation of $\delta_{de}$, Eq.\ (\ref{eq:de-secondorder}), we can see that the effective source term at high $k$ 
(on sub-horizon scales) is proportional to
\be
k^2 \left[ (1+w)\psi + \frac{\delta P_{de}}{\rho_{de}} -\frac{2}{3}\pi_{de} \right]
\approx  k^2 \left[  c_s^2 \Delta_{de} -\frac{2}{3}\pi_{de} \right] \, . \label{eq:effsound} \label{eq:pheno1} 
\ee
Here we neglected the velocity contribution $\propto v/k$ and the potential $\psi$, as both are suppressed by inverse powers of $k$
relative to $\Delta$. We then have a second-order equation for $\Delta_{de}$ with the above term proportional to $\Delta_{de}$. 
If the pre-factor of $\Delta_{de}$ in this expression is positive, then it will lead to an oscillatory behaviour of $\Delta_{de}$ and the behaviour of the dark energy perturbations on small scales will be stable. In this case Eq.\ (\ref{eq:effsound}) can be used to define an effective sound speed for the dark energy. If on the other hand the pre-factor is negative then we expect rapid growth of the perturbations on small scales which in general renders the model unviable. 

Based on these considerations it makes thus sense to define an effective sound speed, which for the models described in the last section takes the form 
\be 
\ceff^2 \equiv c_s^2 -\frac{2 \ff}{3} \, .
\label{eq:pheno2}
\ee
It is this effective sound speed that characterises the propagation of perturbations and the pressure support (and hence the clustering properties) on small scales (see
also \cite{Sapone:2012nh,Sawicki:2012re,Sapone:2013wda} where the same combination was found to be relevant). Here we
also assumed that the scales of interest satisfy $ k^2/\H^2 \mug \gg^2 $.

Since the full system of differential equations cannot be solved analytically in general, we will focus in the next subsections on limiting cases for which dark matter and dark energy perturbations decouple from each other. In some of them we compare our results with the full numerical solutions explicitly, however we have checked for all of them that the approximate expressions show a behaviour that is representative of the full numerical solution in the relevant regime (see Fig. \ref{fig:comparison}).
We found it to be convenient to solve the $4$-dimensional system (\ref{eq:de-cont})-(\ref{eq:de-eul}) for the dark matter and dark energy perturbations by using the dimensionless variables  
\be
V_m \equiv  -\frac{k v_m}{\H}\,, \quad
V_{de} \equiv -\frac{k v_{de} (1+w)}{\H} \,,
\label{eq:pheno3} 
\ee
along with the density contrast variables $ \delta_m$ and $ \delta_{de}$. This choice makes it simpler to expand the equations consistently in powers of $k$ 
to study separately the super- and sub-horizon behaviour, and we checked that we are
are able to recover the solutions for dark matter and dark energy perturbations in the matter dominated era found in \cite{Sapone:2009kx}. 

\subsection{Sub-horizon scales}\label{subsection:4.1}

On sub-horizon scales, $ k/\H \mug  1 $, we find three scenarios where dark matter and dark energy perturbations decouple. These correspond to: i) dark matter domination ii) dark energy domination without dark matter contribution to the dark energy anisotropic stress $ (\aa = 0) $ and iii) the particular case where $ \ff = -1/2 $. Although for the last case we find analytical solutions for dark matter perturbations, they do not seem to have a special physical relevance and we will not discuss this case further.

\subsubsection{Dark matter domination} \label{sec:md_subh}

During dark matter domination the evolution of the conformal Hubble parameter and (neglecting decaying modes and focusing
on sub-horizon scales) the solutions for matter perturbations are given by (e.g.\ \cite{Sapone:2009kx})
\be
\H^2 = H_0^2 \frac{\Omega_m} {a}\, , \qquad \delta_m = V_m = \delta_0 a 
\label{eq:pheno4}  \label{eq:pheno5}
\ee
where  $ \delta_0 $ is a constant. Using the solutions (\ref{eq:pheno5}) it is possible to find a second order equation for the dark energy density perturbations (assuming that $ k^2/\H^2 \mug 9(1+w)/4\aa a^n $) during matter domination which we can write as
\begin{align}
&\delta''_{de}  +  \[ \frac{3 - 6 w + 4 \ff}{2 a} \] \delta'_{de}
+ \[  \frac{9 H_0^2 \Omega_m (1 - 6 \ceff^2 ) (\ceff^2 + \frac{2 \ff}{3} - w)  + 4 \ff \gg^2 H_0^2 \Omega_m + 6 a \ceff^2 k^2 }{6 a^2 H_0^2 \Omega_m}  \] \delta_{de}  
\nonumber \\ 
& \hspace{3cm}
= \frac{2 \delta_0 \aa a^n k^2}{3 H_0^2 \Omega_m}\,. \quad \,
\label{eq:pheno7}
\end{align}
In Principle, this equation can be solved analytically in terms of Bessel and hypergeometric functions (see Eq.\ (\ref{eq:appendix:A1}) of Appendix \ref{appendix:2}). The argument of the Bessel functions is proportional to $\sqrt{\ceff^2}$,  and as in the case of dark energy domination in Sec.\ \ref{subsubsection:3:1:2}, the perturbations grow exponentially fast for $\ceff^2<0$ because the argument of the Bessel functions becomes imaginary. It is however more instructive to look separately at super- and sub-sound horizon limits where we can simplify the equation further and so obtain more tractable solutions. \\

\noindent\textbf{Super-sound horizon (but sub-horizon)}\\

The sound horizon is set by $\H/\ceff$, i.e.\ a given $k$ is super-sound horizon but sub-horizon if $\H\ll k\ll \H/\ceff$. So for a clean separation of scales we need $\ceff \ll 1$, which means we can just take the limit $\ceff\to 0$  in Eq.\ (\ref{eq:pheno7}). We notice from Eq.\ (\ref{eq:pheno2}) that if $ 0\leq c_s^2 \leq 1 $, then $ 0 \leq \ff \leq 3/2 $.\footnote{If $c_s^2$ can take negative values then $ \ff $ can be negative as well since $ \ff = 3c_s^2/2 $ when $\ceff = 0$. This is not a problem for the stability of the perturbations, as that is governed by $\ceff$.} We find
\be 
\delta''_{de} & + & \[ \frac{3 - 6 w + 4 \ff}{2 a} \] \delta'_{de} + \[ \frac{3 (2 \ff - 3 w)  + 4 \ff \gg^2 }{6 a^2}  \] \delta_{de}  = \frac{2 \delta_0 \aa a^n k^2}{3 H_0^2 \Omega_m}\,. 
\label{eq:pheno8}
\ee

The homogeneous part of the equation clearly has power-law solutions, in general the solution for Eq.\ (\ref{eq:pheno8}) is of the form
\be 
\delta_{de} = A_1 a^{\frac{1-\alpha-\beta}{2}} + B_1 a^{\frac{1-\alpha+\beta}{2}} +  \frac{2 \delta_0 \aa k^2 a^{2+n}}{3 H_0^2 \Omega_m [2(1+\alpha) + \vartheta + n (3+\alpha +n)]  }
\label{eq:pheno9}
\ee
where 
\begin{eqnarray}
\vartheta & = & \frac{3 (2 \ff - 3 w)  + 4 \ff \gg^2 }{6}   \\
\alpha & = & \frac{3 - 6 w + 4 \ff}{2}\\ 
\beta & = &  \sqrt{1-2\alpha + \alpha^2 -4 \vartheta }
\label{eq:pheno10}
\end{eqnarray}
and $ A_1 $ and $ B_1 $ are two constants of integration. The last term in Eq.\ (\ref{eq:pheno9}) is a growing mode driven purely by the external anisotropic stress (the part of the anisotropic stress coupled to $\Delta_m$), and it can more clearly be written as
\be
\delta_{de}^{(\aa)} = 
\aa \delta_m \left(\frac{k^2}{\H^2} \right) \frac{2 a^n}{3 [2(1+\alpha) + \vartheta + n (3+\alpha + n)]  } \, .
\ee
The factor $(k/\H)^2$ in this expression interpolates between $1$ on horizon scales (where $k=\H$) and $1/\ceff^2$ on sound horizon scales (where $\ceff k = \H$), i.e.\ between the terms containing $\aa$ in Eqs.\ (\ref{eq:pheno24}) and (\ref{eq:pheno12}).

\begin{figure}[tb]
\centering
\includegraphics[width=.9\textwidth]{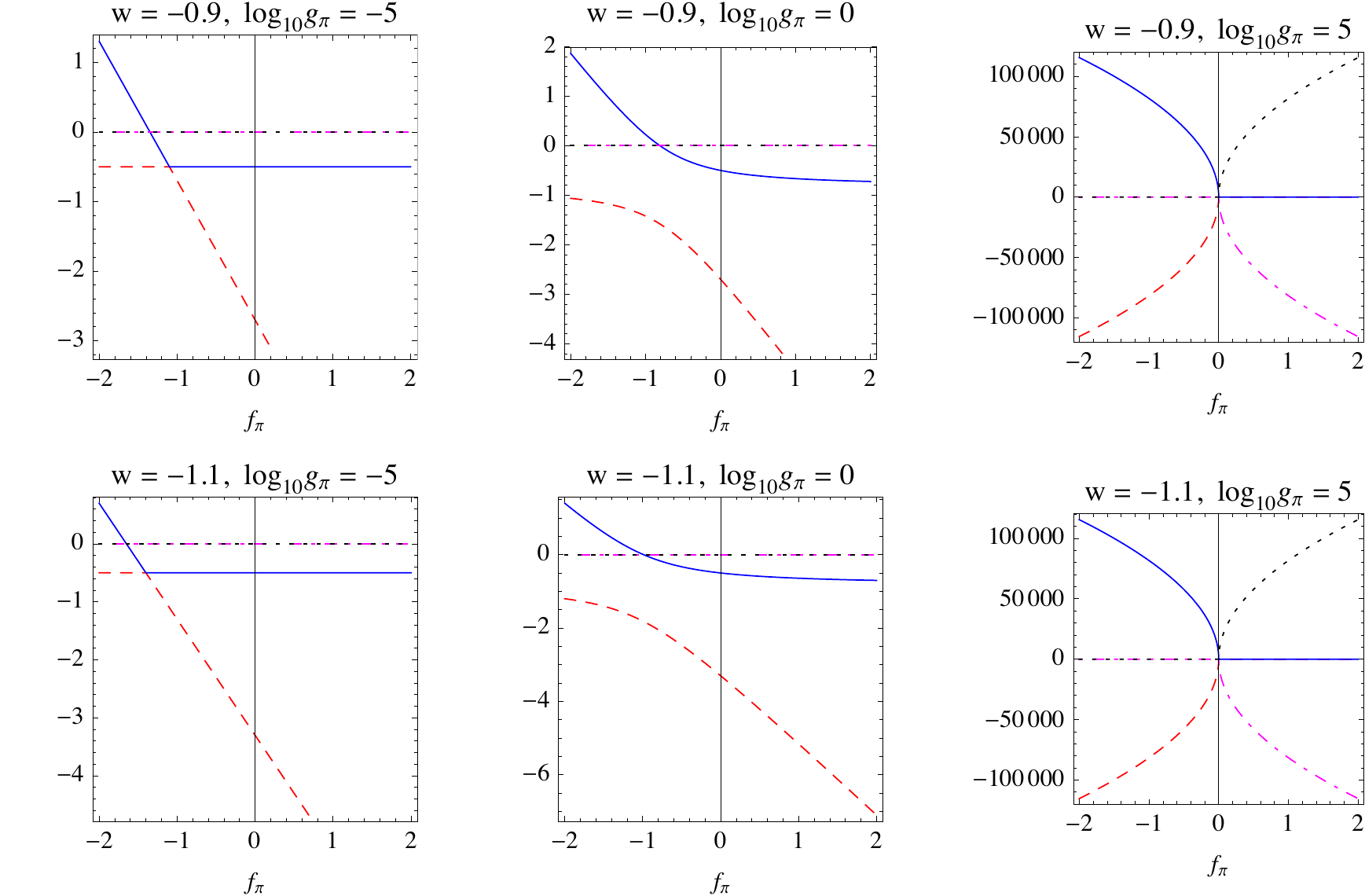}
\caption{The dark energy perturbations during matter domination in the sub-horizon but super-sound horizon regime have a power-law behaviour. Here we show the exponents of the first two terms in Eq. (\ref{eq:pheno9}) for a range of values of the parameters $\ff$ and $\gg$ (the parameter $\aa$ leads to an additional growing mode
driven by the dark matter). Real and imaginary parts for the exponent of the first term are plotted in red dashed and magenta dot-dashed lines, respectively. For the second term the real part is plotted in blue and the imaginary part is shown in black dotted lines. Positive real parts correspond to growing modes, not necessarily instabilities.}
\label{fig:expogrid1}
\end{figure}

We show the exponents of the homogeneous solutions as a function of the parameters $\ff$ and $\gg$ in Fig.\ \ref{fig:expogrid1}. For $\gg \lesssim 1$ we find a growing mode when $\ff \lesssim -1$, and for very negative $\ff$ this mode can grow very quickly. This rapid perturbation growth will eventually lead to a conflict with observations so that we expect to find a lower limit for $\ff$ around $\ff \simeq -3$ to $-5$ based on the growth of dark energy perturbations during dark matter domination. For $\gg \mug 1$ the dark energy perturbations grow extremely fast as soon as $\ff$ becomes negative, rendering this part of parameter space unviable. (For $\ff>0$ the exponent is purely imaginary, so that the dark energy perturbations oscillate without growing.) We will see in the next section that this $\gg$ dependent lower limit on $\ff$ is indeed clearly visible.

With Eq.\ (\ref{eq:pheno9}) we can also find an expression for the dark energy velocity perturbation,   
\begin{eqnarray} 
V_{de} & = & \frac{[B_1 (1+ 4\ff - 6w - \alpha + \beta) a^\beta + A_1 (1+4\ff -6w-\alpha - \beta) ] a^{\frac{1-\alpha - \beta}{2}} }{2} \nonumber \\
&&+ \frac{2 \aa k^2 \delta_0 (2+2\ff -3w + n)a^{2+n}}{3 H_0^2 (2+\vartheta + 2\alpha +n(3+n+\alpha))\Omega_m} \, .
\label{eq:pheno9b}
\end{eqnarray}

\noindent\textbf{Sub-sound horizon}\\

The equations for dark energy velocity and density perturbations in this case are given by
\begin{eqnarray}
V_{de}' + \frac{1-6\ceff^2}{2a} V_{de} &=& -  \frac{\ceff^2 k^2}{H_0^2 \Omega_m}   \delta_{de} 
+ \frac{\delta_0}{6 H_0^2\Omega_m} \[ 4 \aa a^{n+1} k^2 + 9 H_0^2(1+w)\Omega_m\] \, , \label{eq:pheno11} \\
\delta_{de}' + \frac{3(c_s^2-w)}{a}\delta_{de}  &=& \frac{V_{de}}{a} \, . \label{eq:pheno11b}
\end{eqnarray}
Here we re-introduced the second, sub-dominant term in Eq.\ (\ref{eq:pheno11})  for the special case $\aa=0$.
Like \cite{Sapone:2009kx} we can argue that if we want to avoid large velocity perturbations, then we expect the source terms in Eq.\ (\ref{eq:pheno11}) to cancel to a high degree. It follows that 
\begin{eqnarray} 
\label{eq:pheno12}
\delta_{de} 
&=& a \delta_0 \[ \frac{2 \aa a^n}{3 \ceff^2} + \frac{3 H_0^2 \Omega_m (1+w)}{2 a \ceff^2 k^2} \] \\
V_{de} &=& a \delta_0 \[  \frac{2}{3\ceff^2} \aa a^n \left\{1 + 3 (\ceff^2-w) +2\ff +n \right\} \right. \nonumber \\ 
&&+ \left. \frac{3 H_0^2 \Omega_m \left\{3 (\ceff^2-w) +2\ff \right\}(1+w) }{2 a \ceff^2 k^2 } \] \label{eq:pheno13}
\end{eqnarray}
where the last term in each equation is only relevant if $\aa = 0$. We see that during matter domination
the dark energy perturbations in the sub-sound horizon regime only grow if the coupling to $\Delta_m$ is non-zero.
In that case $\delta_{de}$ is proportional to $a^n \delta_m$. If $e_\pi=0$ then the dark energy perturbations become constant on
sub-sound horizon scales in matter domination. However, it should be mentioned that we have here neglected modes that are 
usually decaying (as in appendix B of \cite{Sapone:2009kx}). As
mentioned at the start of the section, if $\ceff^2<0$ then the full solution of Eq.\ (\ref{eq:pheno7}) grows exponentially.

\subsubsection{Dark energy domination and $ \aa = 0 $}\label{subsubsection:3:1:2}

Considering that during dark energy domination the conformal Hubble parameter can be approximated by

\be 
\H^2 = H_0^2 \frac{\Omega_x} {a^{1 + 3 w}}\, ,
\ee 
we find a homogeneous second order equation for the dark energy density perturbations,
\begin{eqnarray}
\label{eq:pheno14}
\delta''_{de} &+&  \bigg\{\,  \frac{3 + 4 \ff - 9 w}{2 a}  \bigg\}\,   \delta'_{de}
+  \bigg\{\, \frac{a^{1 + 3 w} \ceff^2 k^2}{a^2 H_0^2 \Omega_x}  + \frac{\ff ( 2 \gg^2 - 9 - 27 w ) }{3a^2}
 \\ \nonumber
 &-& \frac{3 ( \ceff^2 + \frac{2 \ff}{3} ) \left[6 ( \ceff^2 + \frac{2 \ff}{3})  -  (1 + 4 \ff + 3 w) \right] + 3 (1 - w)(1 + 3 w) }{2 a^2}  \,\bigg\}
 \delta_{de} = 0 \, .
\end{eqnarray} 

Again we can look at both super and sub-sound horizon limits for this equation. \\

\noindent\textbf{Super-sound horizon}\\

\begin{figure}[tb]
\centering
\includegraphics[width=.9\textwidth]{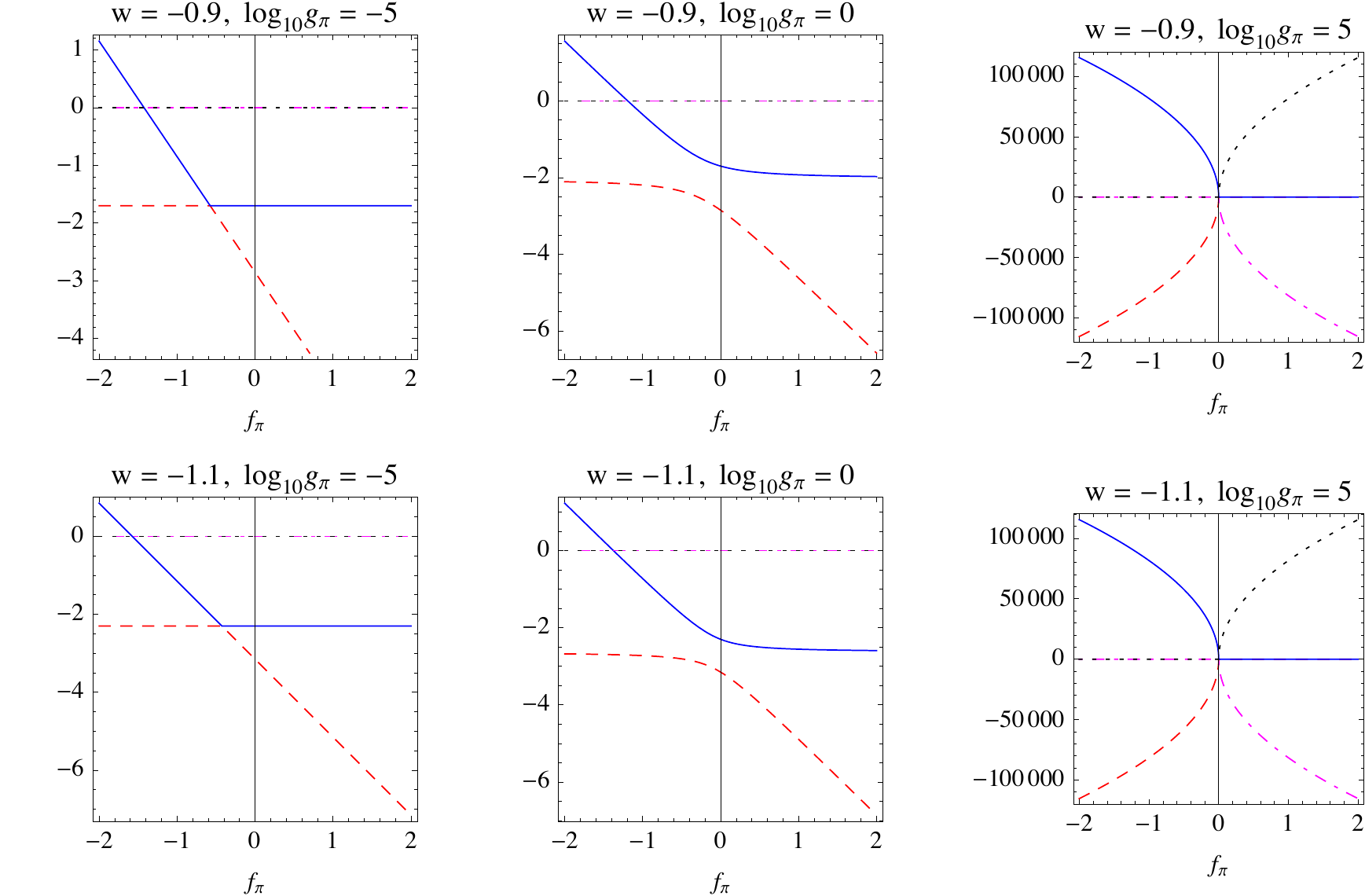}
\caption{The dark energy perturbations during dark energy domination in the sub-horizon but super-sound horizon regime have a power-law behaviour,
with the exponents given here for a range of values of the parameters $\ff$ and $\gg$. Here we plot the exponents of the two terms in Eq. (\ref{eq:pheno16}): red dashed (real part) and magenta dot-dashed (imaginary part) lines correspond to the first term whereas blue (real part) and black dotted (imaginary part) lines correspond to the second term.}
\label{fig:expogrid3}
\end{figure}

In the super-sound horizon limit Eq.\ (\ref{eq:pheno14}) becomes
\be 
\delta''_{de} &+&  \[ \frac{3 + 4 \ff - 9 w}{2 a}  \] \delta'_{de} +  \[  \frac{4 \ff (-3 + \gg^2 - 9 w) + 9 ( 3 w^2 - 2 w - 1 )  }{6 a^2}  \] \delta_{de} = 0 
\label{eq:pheno15}
\ee

which again has power-law solutions given by 
\be 
\delta_{de} = A_3 a^{\frac{1-\alpha_3 - \beta_3}{2}} + B_3 a^{\frac{1-\alpha_3 + \beta_3}{2}}
\label{eq:pheno16}
\ee
where 
\begin{eqnarray}
\label{eq:pheno17}
\alpha_3 & = & \frac{3+4\ff-9 w}{2} \nonumber \\
\vartheta_3 & = & \frac{4\ff (-3+\gg^2 -9 w) + 9 (3 w^2 - 2 w -1) }{6}\nonumber \\
\beta_3 & = & \sqrt{1-2 \alpha_3 + \alpha_3^2 - 4 \vartheta_3}
\end{eqnarray}
 We plot the behaviour of the exponents in Fig.\ \ref{fig:expogrid3}. Overall, the behaviour is similar to the one shown in Fig.\ \ref{fig:expogrid1}: For small $\gg$ the perturbations can grow rapidly if $\ff \lesssim -3$ while for large $\gg$ they grow quickly whenever $\ff<0$. 
For velocity and matter perturbations we have
\begin{eqnarray} 
\label{eq:pheno16b}
V_{de} &=& \frac{a^{\frac{1-\alpha_3 - \beta_3}{2}}}{2}  \[  B_3 a^{\beta_3} (1+6c_s^2 -6w - \alpha_3 + \beta_3)  + A_3 ( 1+6c_s^2 -6w - \alpha_3 - \beta_3 ) \] \nonumber \\
\delta_m &=& 6(1+2\ff)a^{\frac{1-\alpha_3 - \beta_3}{2}} \[ \frac{B_3 a^{\beta_3}}{(1-\alpha_3 +\beta_3)(2-3w-\alpha_3 + \beta_3)}  \right. \nonumber \\
&& + \left.  \frac{A_3}{(-1+\alpha_3 + \beta_3)(-2+3w+\alpha_3 + \beta_3)}\] + \delta_0 \nonumber \\
V_m &=& \frac{3(1+2\ff) a^{\frac{1-\alpha_3 - \beta_3}{2}} \[ A_3 (-2+3w+\alpha_3 - \beta_3)  + B_3 a^{\beta_3} (-2+3w+\alpha_3 + \beta_3) \] }{(2-3w-\alpha_3 +\beta_3)(-2+3w+\alpha_3 + \beta_3)}
\end{eqnarray}
where we have neglected a decaying mode in the matter density perturbation. We can see that the dark matter density perturbation $\delta_m$ follows the dark energy perturbations and grows at the same rate (in addition to a constant mode). \\

\noindent\textbf{Sub-sound horizon}\\

On the other hand, in the sub-sound horizon limit and if we assume 
\be 
\frac{\ceff^2 k^2}{\H^2} \mug \frac{\ff ( 2 \gg^2 - 9 - 27 w ) }{3} - \frac{3 c_{s}^2 \[6 c_{s}^2  -  (1 + 4 \ff + 3 w) \] + 3 (1 - w)(1 + 3 w) }{2}\nonumber
\ee
equation (\ref{eq:pheno14}) reads
\be 
\delta''_{de} &+&  \[ \frac{3 + 4 \ff - 9 w}{2 a}  \] \delta'_{de} +  \[  \frac{\ceff^2 k^2}{ H_0^2 \Omega_x a^{1-3w}}  \] \delta_{de} = 0 
\label{eq:pheno18}
\ee
and we expect to have exponential growth if $ \ceff^2<0 $. The general solution of Eq.\ (\ref{eq:pheno18}) is given by
\begin{eqnarray}
\label{eq:pheno19}
\delta_{de} & = & \(\frac{x_3}{2}\)^{\frac{1-\alpha_4}{1+3w}} \lcb A_4\, J_{\nu_1}(x_3) + B_4\, J_{-\nu_1}(x_3)   \rcb 
\end{eqnarray}
where 
\begin{eqnarray}
\label{eq:pheno20}
\alpha_4 & = & \frac{3+4\ff -9w}{2} \nonumber \\
\nu_1 & = & \frac{\alpha_4 -1}{1+3w} \nonumber \\
x_3 & = & \frac{2 a^{\frac{1+3w}{2}} \ceff k}{(1+3w) H_0\sqrt{\Omega_x}}\, ,
\end{eqnarray}
and $ A_4 $ and $ B_4 $ are constants.\footnote{Gamma functions, similar to those appearing in solutions in Appendix \ref{appendix:2},  have been absorbed in the constants $ A_4 $ and $ B_4 $. These constants are fixed by the initial conditions.} We see that for $\ceff^2<0$ the argument $x_3$ of the Bessel functions becomes imaginary, and indeed the perturbations will grow exponentially. Stable perturbations in this regime thus require $\ff < 3 c_s^2/2$. We can also see that the overall pre-factor of Eq.\ (\ref{eq:pheno19}) behaves like $a^{(1-\alpha_4)/2}$, where the exponent is linearly decreasing with $\ff$, i.e.\ the dark energy perturbations grow faster for more negative $\ff$. We therefore expect also a lower cutoff for $\ff$, around $\ff \approx -7$.

\subsection{Super-horizon scales} 

When considering super-horizon scales, $ k/\H \ll 1 $, we find that dark matter and dark energy perturbations decouple from each other again in matter and dark energy domination. 
However, we could only find analytical solutions during matter dominance.

\subsubsection{Dark matter domination}

Since scales larger than the horizon are also super-sound horizon scales, we can set $ \ceff=0 $ which according to Eq.\ (\ref{eq:pheno2}) is equivalent to setting $ c_s^2 = 2\ff/3 $. Then,  if we use Eq.\ (\ref{eq:pheno4}) for the Hubble parameter and neglect decaying modes, we find the following set of solutions for matter and dark energy perturbations
\begin{eqnarray} 
V_m &=& \delta_0 a \, , 
\qquad 
\delta_m = \delta_0 \frac{3 H_0^2 \Omega_m}{k^2} \, , 
\label{eq:pheno21}
\label{eq:pheno22} \\
V_{de} &=& \delta_0 a \[ \frac{ 4 \aa a^n}{4\ff-3-2n} + \frac{3(1+w)}{ 3-4\ff } \] \,,  
\label{eq:pheno23} \\
\delta_{de}& =&  \delta_0  \frac{3 H_0^2 \Omega_m}{k^2 } \[   \frac{4 \aa a^n (2\ff -3w)}{(4\ff-3-2n)(2\ff +n -3w)} - \frac{3(1+w) }{4\ff -3 }\] \, .
\label{eq:pheno24}
\end{eqnarray} 
We see that outside of the horizon the dark energy density perturbation grows like $a^n$ -- in the particular case when $ n=0 $ or $\aa=0$, the dark energy density perturbation (like the dark matter one) is always constant on super-horizon scales. We also notice that it 
 is non-zero only if either the dark energy is coupled to the dark matter through $\aa\neq0$ or if $w\neq-1$ \footnote{But notice that on sub-horizon scales a non-zero anisotropic stress of the dark energy itself can drive the dark energy perturbations even if $w=-1$ and $\aa=0$, see section \ref{sec:md_subh}. However, for our model 2 the perturbations only grow if $\ff<-5/4$ and only in the sub-horizon but super-sound horizon regime.}.
We also note that we recover the solutions found in \cite{Sapone:2009kx} in the absence of anisotropic stress.

\begin{table}[h!]
\centering
\begin{tabular}{|c|c|c|c|}
\hline 
\multicolumn{2}{|c|}{scales} & \multicolumn{2}{|c|}{rapid growth} \\ \cline{3-4}   
\multicolumn{2}{|c|}{} & matter dominance & dark energy dominance ($\aa=0$) \\ \cline{1-4}
\multicolumn{1}{|c|}{\multirow{2}{*}{sub-horizon}} & \multirow{1}{*}{sub-sound}  & $ \phantom{\Bigg|} \ceff^2<0 \phantom{\Bigg|}$  &  $\displaystyle \ceff^2<0 $ \\ \cline{2-4}
 & \multirow{1}{*}{super-sound} & $\phantom{\Bigg|} \ff \ll \dfrac{9w}{2(3+2\gg^2)} \phantom{\Bigg|}$ & $\ff \ll \dfrac{27w^2 -18w -9}{4(3-\gg^2+9w)} $ \\ \cline{1-4}
\end{tabular}
\caption{Regimes and regions in parameter space where dark energy perturbations grow rapidly.}
\label{tab:stability}
\end{table}

We summarise in Tab.\ \ref{tab:stability} the regions in parameter space where we expect rapid growth of the perturbations
that is not compatible with the existence of a stable universe. The growth of the perturbations on sub-sound horizon scales for $\ceff^2<0$
is exponential and corresponds to the usual instability for negative sound speeds. For theories with a given $c_s^2$ this provides an
upper limit for the parameter $\ff$, namely $\ff < 3 c_s^2 /2$. On scales that are sub-horizon but lie above the sound horizon,
the perturbations grow as a power law with a very high power for sufficiently negative $\ff$, as indicated in the table.\footnote{The regions in the table correspond to the sub and super-sound horizon limits of the second order Eqs. (\ref{eq:pheno7}) and (\ref{eq:pheno14}). Note that according to the general solution for matter dominance and sub-horizon scales, Eq. (\ref{eq:appendix:A1}),  there is also exponential growth on super-sound horizon scales for $ \ceff^2<0 $.} This provides
a lower limit for $\ff$ as such a rapid growth of the dark energy is again not compatible with the data.

\begin{figure}[tb]
\centering
\includegraphics[scale=.76]{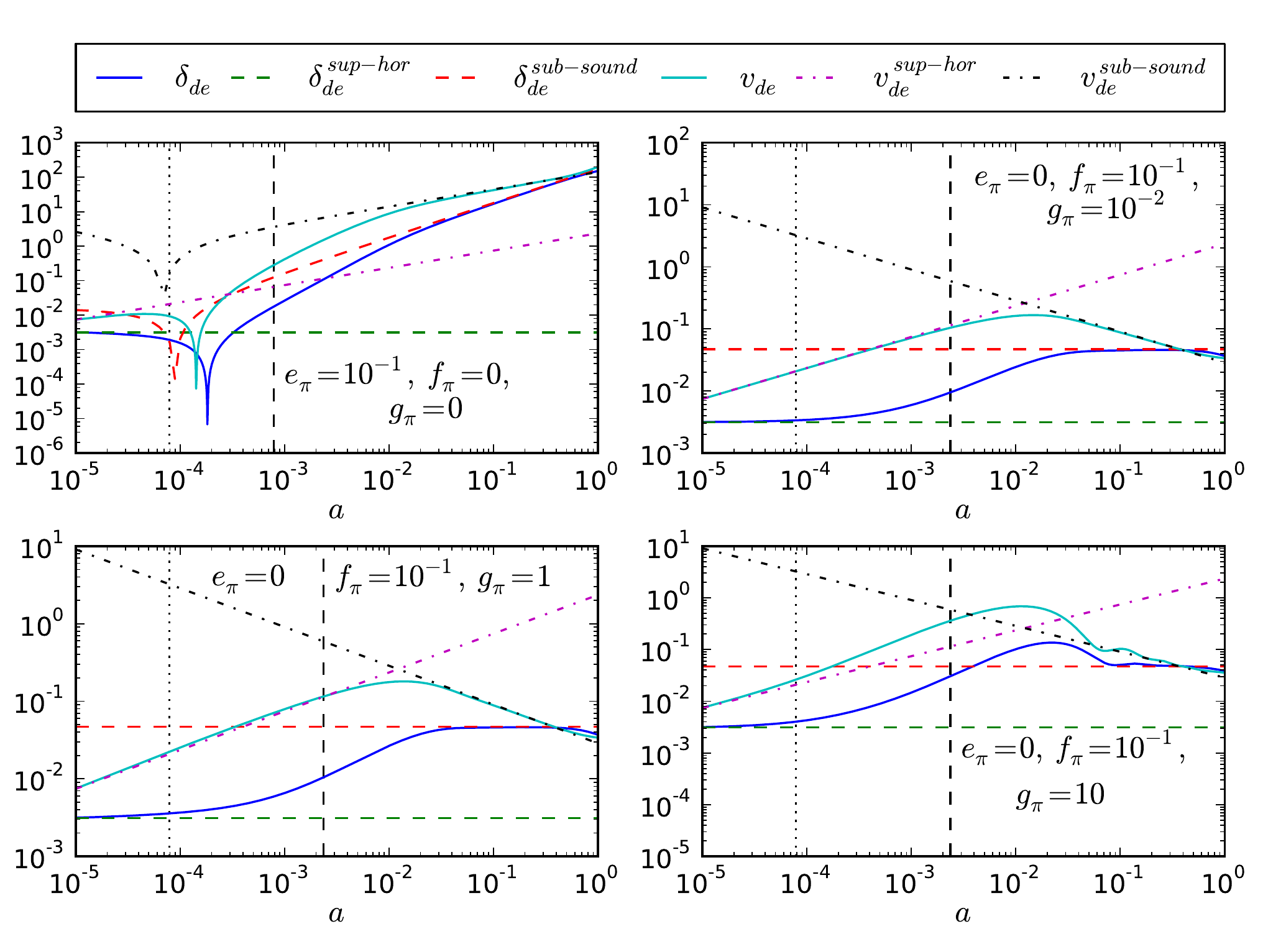} 
\caption{The figure shows the behaviour of the variables $ \delta_{de} $ and $ v_{de} $ for $k=1.5\times 10^{-2}$, $c_s^2 = 10^{-1}$, $w=-1.05$, $ n=0 $ (the power in Eq. (\ref{eq:model:1})) and different combinations of parameters $\aa$, $\ff$ and $\gg$. The blue and cyan curves are the numerical solutions. Green dashed and magenta dot-dashed curves are analytical solutions on super horizon scales, Eqs. (\ref{eq:pheno23})-(\ref{eq:pheno24}). On the other hand, the red dashed and black dot-dashed curves are analytical solutions on sub-sound horizon scales given by Eqs. (\ref{eq:pheno12})-(\ref{eq:pheno13}). The vertical lines give the scale factor at which the mode enters the effective sound horizon (dashed line) and the Hubble horizon (dotted line). We consider  a longer dynamic range in $a$ to illustrate the transition from super-horizon to sub-sound horizon scales more clearly, without however including radiation in the numerical solution.}
\label{fig:comparison}
\end{figure}

In the next section we are going to vary $\ceff>0$ and $\ff$ independently, so that $c_s^2$ can take any value. For this reason we
will not see the upper cutoff on $\ff$ from the instability arising due to $\ceff^2<0$, as we never enter in this regime, but we will
see the lower cutoff. Also, as in the approximate solutions shown in Fig.\ \ref{fig:comparison}, we will limit ourselves to $n=0$ for
the model 1 defined by Eq.\ (\ref{eq:model:1}).

\section{Observational constraints}\label{section:4}

In this section, we investigate the parameter degeneracies and the constraints on the anisotropic stress models for dark energy imposed by different cosmological observations. We use a modified version of the {\tt CosmoMC} code (version Mar 13 \cite{Lewis:2002ah,Cosmomc}) to perform Markov-chain Monte-Carlo explorations of the model likelihoods. The sampler calls a modified version of the {\tt CAMB} code\footnote{The modified codes are available at \href{http://cosmology.unige.ch/content/cosmomc-and-camb-early-dark-energy-and-anisotropic-stress}{http://cosmology.unige.ch/content/cosmomc-and-camb-early-dark-energy-and-anisotropic-stress} and the chains (1.4\,GB) can be downloaded from \href{http://theory.physics.unige.ch/~kunz/traces-anisotropic-dark-energy/ade_chains.tar.gz}{http://theory.physics.unige.ch/$ \sim $kunz/traces-anisotropic-dark-energy/ade$\_$chains.tar.gz}.} (version Mar 13 \cite{Lewis:1999bs,Camb}) to compute the linear theory CMB spectra for a given model. 
In all cases we use constraints on the number of relativistic degrees of freedom at Big-Bang nucleosynthesis (BBN) \cite{Pisanti:2007hk} and put a prior on the age of the Universe to be between 10 and 20 Gyrs. In addition we use the CMB likelihood code of the Planck collaboration (version 1.0 \cite{Planck:2013kta,Ade:2013zuv}) which includes the Planck first data release combined with WMAP 9yr low-multipole polarisation data \cite{Bennett:2012fp}. Moreover, we add the high-multipole temperature data from the South Pole Telescope (SPT) \cite{Story:2012wx} and the Atacama Cosmology Telescope (ACT) \cite{Das:2013zf}. In the following, we use the abbreviation \emph{CMB data} for the combination of Planck temperature, WMAP9 low-multipole polarisation, and ACT and SPT temperature  (referred to as \emph{Planck}+WP+highL in the Planck papers). We expect the CMB data to provide constraints not only on the parameters that describe the primordial power spectrum and the re-ionisation history but also to mildly constrain the late-time evolution of the gravitational potentials through the integrate Sachs-Wolfe effect and CMB lensing.

To further constrain the parameters that are relevant for the late-time evolution of the background geometry, the density parameter and equation of state of dark energy, we also use constraints on the distance--redshift relation from baryon acoustic oscillations (BAO) and type Ia supernovae (SNe). Currently, there are seven BAO measurements available: two from the Sloan Digital Sky Survey (SDSS) DR7 
\cite{Percival:2009xn,Padmanabhan:2012hf}, one from the 6dF Galaxy Survey 
\cite{Beutler:2011hx}, three from the WiggleZ Dark Energy Survey 
\cite{Blake:2011en}, and one from the SDSS-III Baryon Oscillation Spectroscopic Survey (BOSS) DR9 
\cite{Anderson:2012sa}. In case of the supernovae, we use the compilation of 473 SNe Ia provided by the SuperNova Legacy Survey (SNLS) team \cite{Conley:2011ku}.
The fact that Planck prefers a slightly different value for $H_0$ than local measurements of the Hubble parameter in a flat $\Lambda$CDM cosmology raises concerns about the compatibility of these data sets \cite{Ade:2013zuv}. For this reason we chose not to include constraints on the local expansion rate. If we included
the $H_0$ constraint of \cite{Riess:2011yx} we would find that the confidence intervals for $w$ are shifted slightly towards more
negative values, with $w = -1$ sitting close to the $2 \sigma$
limit. On the other hand, we would not find significant changes in the
constraints on the parameters that govern the dark energy
perturbations when including $H_0$ data.

Using large-scale structure data like the galaxy power spectrum $P(k)$ correctly in the context
of dark energy and modified gravity models is quite involved. There are hidden model assumptions
in the analysis of the data and the construction of the likelihood. For example, the background cosmology
is used when
converting angles and redshifts to $k$ vectors. Moreover, the impact of modifications of  
gravity on galaxy bias and non-linear clustering is mostly unknown. For these reasons we limit ourselves for the time
being to the data sets mentioned above.

For the parameter estimation we vary a base set of seven parameters (those of the flat wCDM model). These are the amplitude, $\ln[10^{10}A_s]$, and the tilt, $n_s$, of the spectrum of primordial scalar curvature perturbations (modelled as a power law normalised at $k=0.05\,{\rm Mpc}^{-1}$), the reionisation optical depth, $\tau$, the physical baryon and cold dark matter energy fractions, $\Omega_b h^2$ and $\Omega_c h^2$, 100 times the ratio of the sound horizon to the angular diameter distance to the last-scattering surface, $\theta$, and finally the constant equation of state parameter of dark energy, $w$. In the figures we will replace the ``fundamental'' parameters $A_s$ and $\theta$ by the variance of fluctuations in spheres of 8 Mpc today, $\sigma_8$, and the value of the Hubble parameter today, $H_0$ (in units of km$/$s/Mpc), that are both derived parameters.

In addition to the base model, we vary or fix the values of the parameters that describe the properties of the dark energy perturbations: the effective sound speed, $\log_{10}\ceff^2$, the external anisotropic stress parameter, $e_\pi$, and the internal anisotropic stress parameter, $f_\pi$, with its transition scale, $\log_{10}g_\pi$. We use flat priors for all parameters, set adiabatic initial conditions for the evolution of the cosmological perturbations, and ignore vector and tensor modes for simplicity.

\begin{figure}[tb]
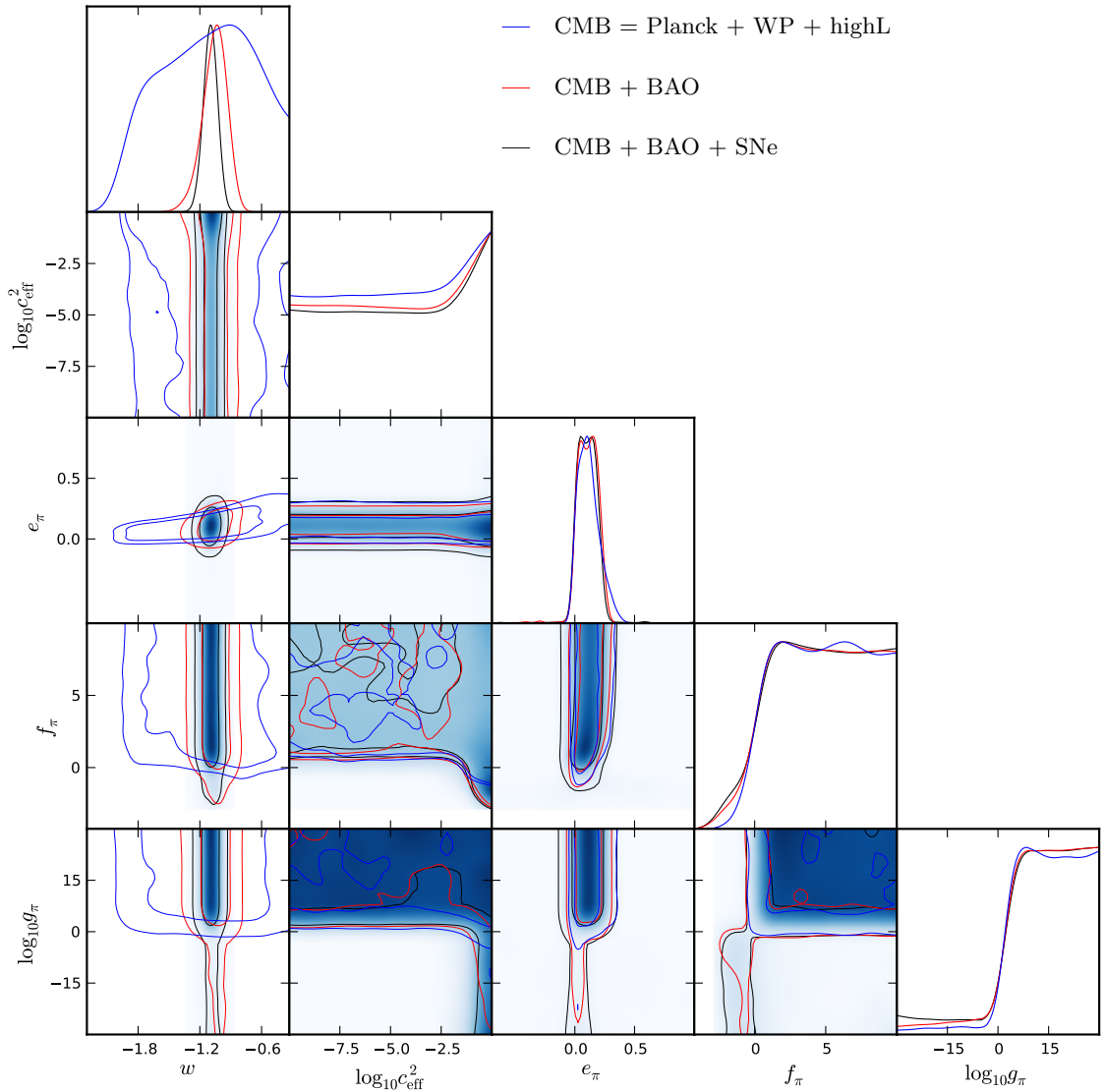

\centering
\begin{lpic}[clean]{PWHiBSwefc_tri(15.5cm,)}
\lbl[l]{90,185; \footnotesize\textcolor{blue}{\bf ---}\quad CMB =
Planck + WP + highL}
\lbl[l]{90,175; \footnotesize\textcolor{red}{\bf ---}\quad CMB + BAO}
\lbl[l]{90,165; \footnotesize\textcolor{black}{\bf ---}\quad CMB + BAO + SNe}
\end{lpic}\\[-0.8cm]
\caption{Marginalised 2d likelihoods and 1- and 2-$\sigma$ contours of combinations of the model parameters $\{w$, $e_\pi$, $f_\pi$, $\log_{10}g_\pi$, $\log_{10}\ceff^2\}$. We compare the use of different data sets: blue is \emph{CMB data} only, red is \emph{CMB+BAO}, and black and the likelihood density plots are \emph{CMB+BAO+SNe}. In all cases we vary all parameters.}
\label{fig:2d_data}
\end{figure}

Let us first take a look at the effect of the different data sets on the parameter constraints. In figure \ref{fig:2d_data} we show the marginalised posteriors and the marginalised 2d-likelihood contours of a parameter subset in the full model, i.e.\ varying all dark energy parameters including the anisotropic stress model, $e_\pi$, $f_\pi$, $g_\pi$ (for $n=0$). We compare the effect of adding more data: blue is \emph{CMB data} only, red is \emph{CMB+BAO}, and black and the likelihood density plots are \emph{CMB+BAO+SNe}. For parameters that are not related to dark energy anisotropic stress the likelihood contours shrink considerably when adding the low-redshift data, as they contain much information on the late-time expansion and therefore on $w$. The constraints on the anisotropic stress parameters are not much altered by adding low-redshift data because BAO and SNe do not contain information on the growth of structure that is affected by the dark energy clustering. To improve those constraints we would need to add information on galaxy clustering, redshift space distortions and cosmic shear.

 The $(\ff,\, \log_{10}g_\pi)$ plane also shows nicely the lower limit on $\ff$ from the rapid growth of perturbations. As argued in the discussions of the sub horizon / super-sound horizon perturbation evolution in  section \ref{subsection:4.1} and shown in  Figs.\ \ref{fig:expogrid1} and \ref{fig:expogrid3}, a very negative value of $\ff$ is in conflict with observations as the dark energy perturbations become large. We can also see how the lower limit on $\ff$ changes as a function of $\gg$, with $\gg \mug 1$ requiring $\ff>0$. 
 
It is interesting to note that the marginalised likelihood in the $(\log_{10}\ceff^2,\, \ff)$ plane peaks where $f_\pi$ is negative and $\log_{10}\ceff^2$ close to 0, while the $(\ff,\, \log_{10}g_\pi)$ plane shows that the likelihood for negative $f_\pi$ is much lower than for positive $\ff$. Note that this is a volume effect of the marginalization. If we were to fix $\ceff=1$, then the high plateau of the likelihood in the $(\ff,\, \log_{10}g_\pi)$ plane would shift from the positive--positive to the negative--negative quadrant, and the situation would be quite different.

Thus, if we had additional observables that even more strongly prefer $\ceff=1$, we would conclude $f_\pi \leq 0$. The constraints on $e_\pi$ would not be affected as it is virtually not degenerate with $\log_{10}\ceff^2$. There is a hint that this could actually be the case: the CFHTLens weak lensing survey \cite{Kilbinger:2012qz} as well as the Planck cluster counts \cite{Ade:2013lmv} prefer $\sigma_8$ and $\Omega_m$ considerably lower than Planck alone (in the $\Lambda$CDM model). We added a toy constraint on the combination $\sigma_8(\Omega_m/0.27)^{0.6}$ that reflects the weak lensing and cluster counts, and noticed that it is passed through the parameter degeneracies in such a way that it constrains $\ceff^2 \sim 1$ and $f_\pi$ and $\log_{10}g_\pi$ both negative. In this case $f_\pi$ turns out to be strongly constrained and $f_\pi=0$ is already in quite some tension with the toy data. We interpret this as a hint that additional dark energy degrees of freedom are able to reconcile apparent tensions between different current data sets. However, we emphasise that the analysis of the weak lensing and cluster count likelihood needs to be done fully correctly within the framework of a generalised dark energy model like ours, and the constraints quoted in the literature \cite{Kilbinger:2012qz, Ade:2013lmv} cannot directly be implemented since they are derived for the $\Lambda$CDM model.

\begin{figure}[tb]
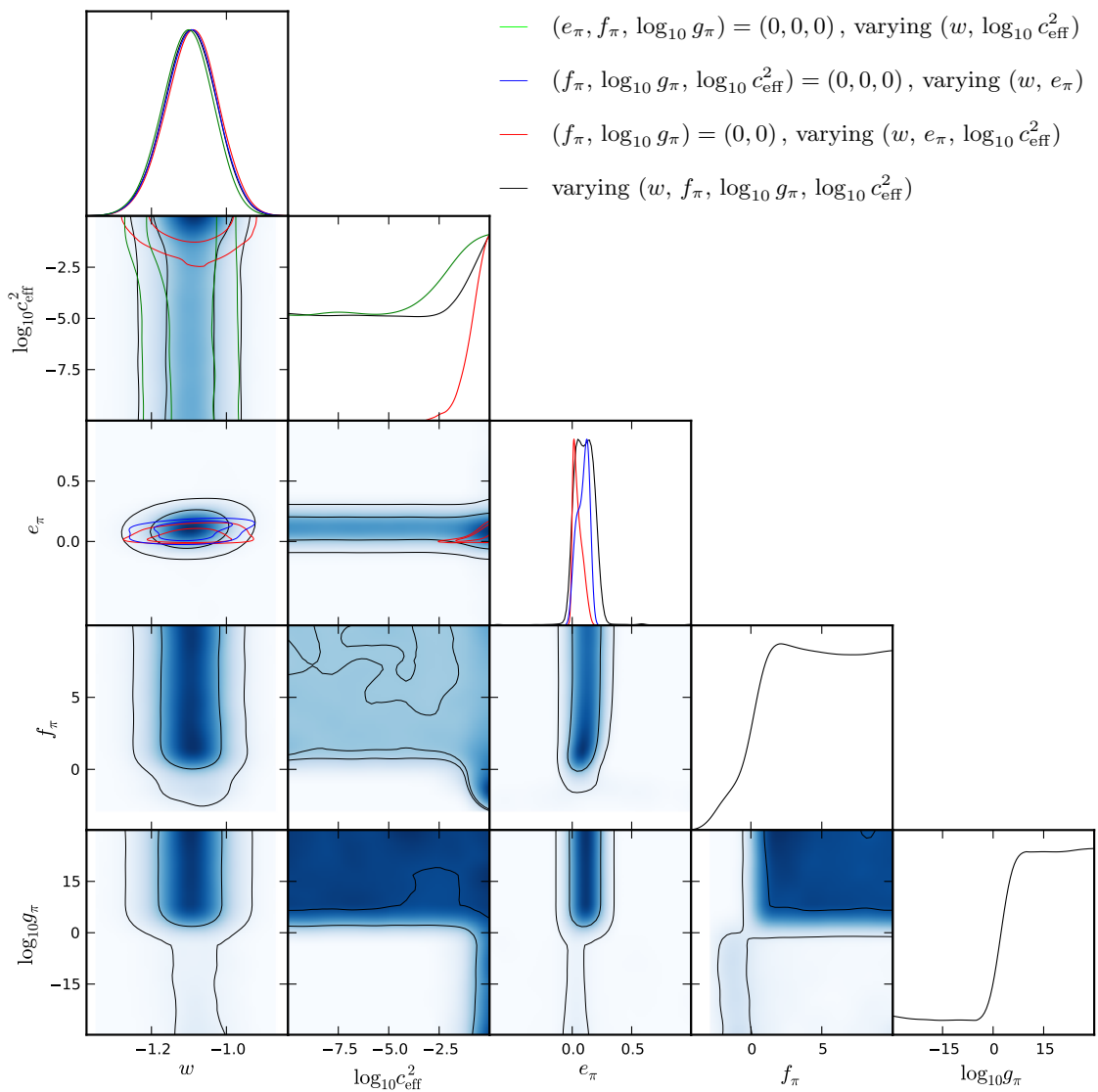

\centering
\begin{lpic}[clean]{PWHiBSwefc_models_tri(15.5cm,)}
\lbl[l]{90,186; \footnotesize\textcolor{green}{\bf ---}\quad
$(\aa,f_\pi,\,\log_{10}g_\pi)=(0,0,0)\,$, varying
$(w,\,\log_{10}\ceff^2)$ }
\lbl[l]{90,177; \footnotesize\textcolor{blue}{\bf ---}\quad
$(f_\pi,\,\log_{10}g_\pi,\,\log_{10}\ceff^2)=(0,0,0)\,$, varying
$(w,\,e_\pi)$}
\lbl[l]{90,168; \footnotesize\textcolor{red}{\bf ---}\quad
$(f_\pi,\,\log_{10}g_\pi)=(0,0)\,$, varying
$(w,\,e_\pi,\,\log_{10}\ceff^2)$}
\lbl[l]{90,159; \footnotesize\textcolor{black}{\bf ---}\quad varying
$(w,\,f_\pi,\,\log_{10}g_\pi,\,\log_{10}\ceff^2)$}
\end{lpic} \\[-0.8cm]
\caption{Marginalised 2d likelihoods and 1- and 2-$\sigma$ contours of combinations of the model parameters $\{n_s$, $\sigma_8$, $w$, $e_\pi$, $f_\pi$, $\log_{10}g_\pi$, $\log_{10}\ceff^2\}$. We compare the different models: 
 for blue we fix $(f_\pi,\,\log_{10}g_\pi,\,\log_{10}\ceff^2)=(0,0,0)$, for red we fix $(f_\pi,\,\log_{10}g_\pi)=(0,0)$, and for black and the likelihood density plots we vary all parameters (except for the scaling exponent $n$ of model 1 which is always set to $n=0$). Here we are using the full data set,  \emph{CMB+BAO+SNe}.}
\label{fig:2d_models}
\end{figure}

\begin{figure}[tb]
\centering
\vspace{2 cm}
\begin{lpic}[clean,nofigure]{fig-6-nolabel(15.5cm,)}
\lbl[l]{10,175; \footnotesize\textcolor{black}{\bf ---}\quad
CMB$ + $BAO$ + $SNe}
\lbl[l]{10,166; \footnotesize\textcolor{red}{\bf ---}\quad
CMB$ + $BAO}
\lbl[l]{10,157; \footnotesize\textcolor{blue}{\bf ---}\quad
CMB$ = $PLANCK$ + $WP$ + $highL}
\lbl[l]{130,180; \footnotesize\textcolor{black}{\bf ---}\quad varying
$\log_{10} \ceff^2,\, \aa,\,f_\pi,\,\log_{10}g_\pi$}
\lbl[l]{130,171; \footnotesize\textcolor{red}{\bf ---}\quad varying
$ \log_{10}\ceff^2,\,\aa $}
\lbl[l]{130,162; \footnotesize\textcolor{blue}{\bf ---}\quad varying
$ \aa $}
\lbl[l]{130,153; \footnotesize\textcolor{green}{\bf ---}\quad varying
$ \log_{10}\ceff^2 $}
\includegraphics[width=.48\textwidth]{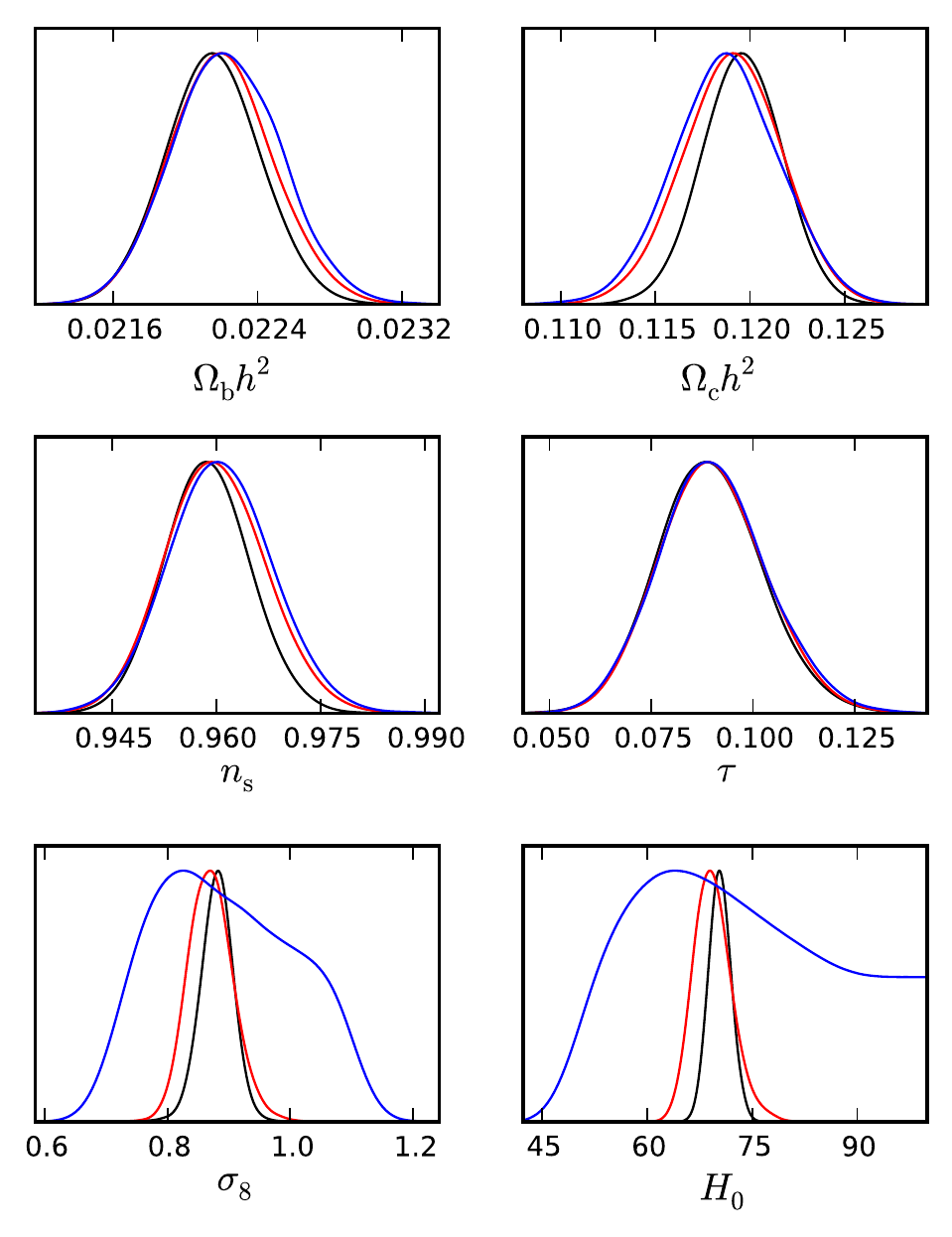}\ 
\includegraphics[width=.50\textwidth]{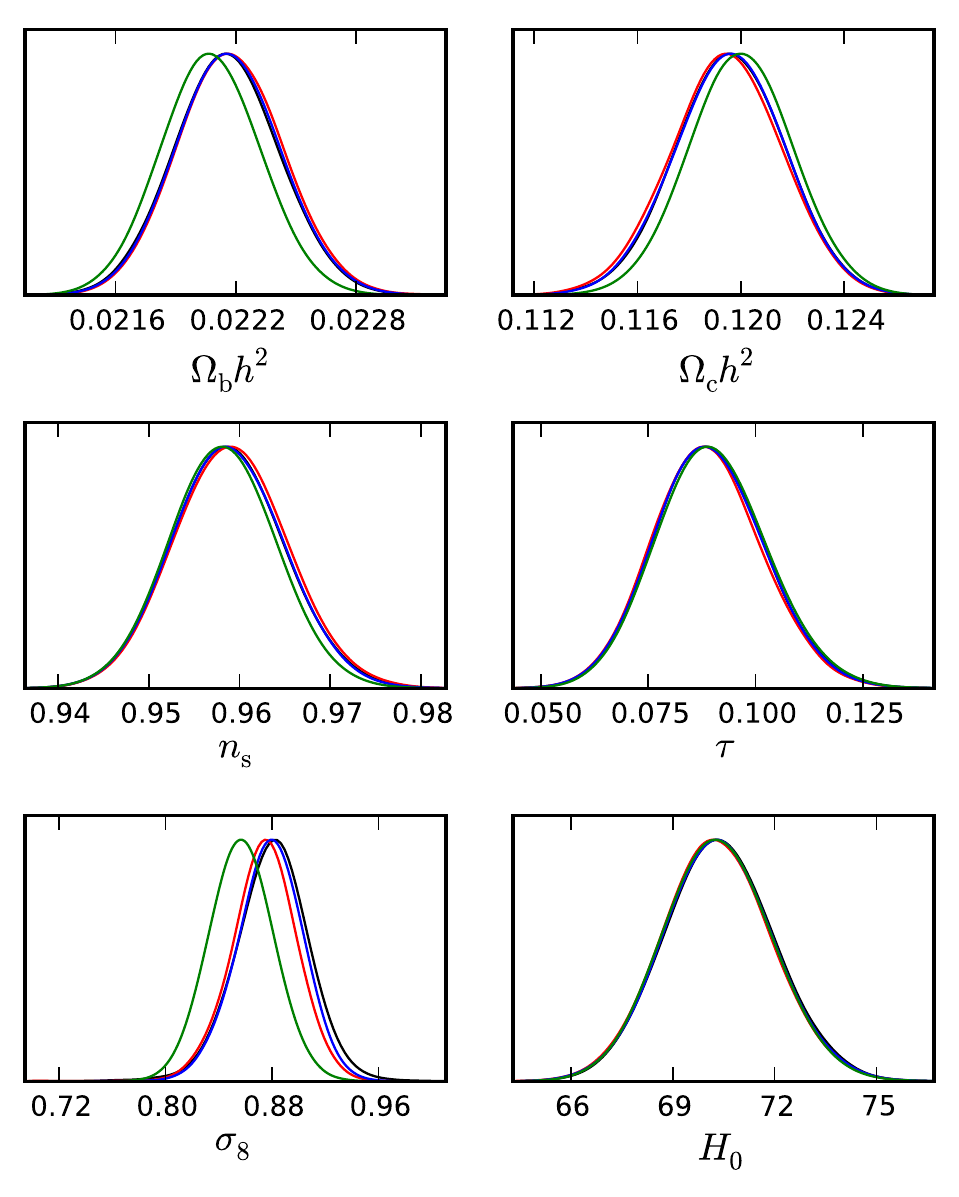}
\end{lpic}\\[-0.1 cm]
\caption{Marginalised posteriors of those model parameters not directly related to dark energy $\{\Omega_bh^2$, $\Omega_ch^2$, $n_s$, $\tau$, $\sigma_8$, $H_0\}$. 
\emph{Left panel}: comparison of different data sets as in figure \ref{fig:2d_data}. The addition of background data sets helps to constrain especially $H_0$ and $\sigma_8$.
\emph{Right panel}: comparison of different models as in figure \ref{fig:2d_models}. The constraints on these parameters do not change significantly when the anisotropic stress is non-zero, with the exception of $\sigma_8$ which prefers a slightly higher value.}
\label{fig:1d_other}
\end{figure}

Next, let us study the anisotropic stress model parameters. In figure \ref{fig:2d_models} we show the marginalised posteriors and 2d-likelihoods using the full data set, \emph{CMB+BAO+SNe}. We compare the different models: for green we fix $(e_\pi,\,f_\pi,\,\log_{10}g_\pi)=(0,0,0)$, for blue we fix $(f_\pi,\,\log_{10}g_\pi,\,\log_{10}\ceff^2)=(0,0,0)$, for red we fix $(f_\pi,\,\log_{10}g_\pi)=(0,0)$, and for black and the likelihood density plots we vary all parameters. We observe that in case of no anisotropic stress, green, the dark energy sound speed is only very mildly preferred to be close to 1, as expected from earlier studies, see e.g.\ \cite{Bean:2003fb,dePutter:2010vy}. Finally, in figure \ref{fig:1d_other} we show the marginalised posteriors for the remaining six base parameters, $\{\Omega_bh^2$, $\Omega_ch^2$, $n_s$, $\tau$, $\sigma_8$, $H_0\}$, that are not directly related to dark energy. We note that all models with dark energy anisotropic stress slightly prefer a higher $\sigma_8$ than in the smooth dark energy case, $\ceff=c_s=1$ (see \cite{Kunz:2003iz} for a study of the impact of $w$ on $\sigma_8$ in smooth dark energy models). This is compatible with the discussion above on the slight tension between constraints on $\sigma_8$ from Planck CMB and Planck cluster counts as well as weak lensing.

\begin{figure}[tb]
\centering
\includegraphics[width=\textwidth]{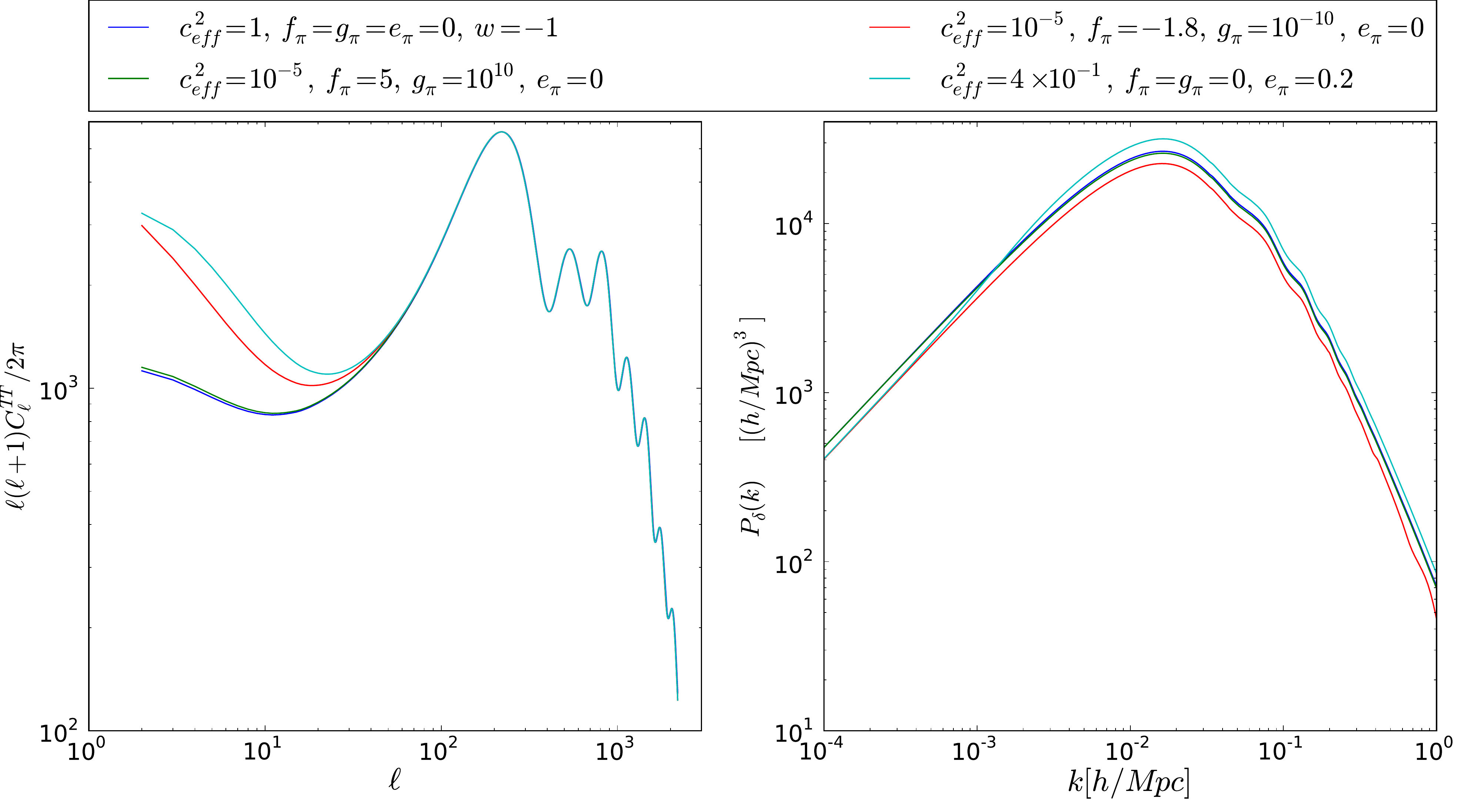}
\caption{CMB angular power spectra (left panel) and matter power spectra (right panel). The concordance model is shown in blue. In green we plot a model with internally sourced anisotropic stress whose parameters are allowed by the cosmological constraints. Two models with parameters excluded by cosmological constraints are depicted in red (internally sourced anisotropic stress) and cyan (externally sourced anisotropic stress). For those models different from the concordance model we used $ w=-0.95 $. In the CMB the differences appear on large scales as the ISW effect is strongly affected by the late-time anisotropic stress of the dark energy. The impact on the matter $P(k)$ is less strong and on scales smaller than the peak appears mostly as a shift in the normalisation (and thus a shift in $\sigma_8$), although this is different on large scales (that are however difficult to observe in galaxy surveys). The effect looks degenerate with an early dark energy contribution (see e.g.\ Fig.\ 5 of \cite{Hollenstein:2009ph}) and it may also be difficult to distinguish observationally from galaxy bias.}
\label{fig:pk-cls}
\end{figure}


\section{Modified growth parametrisations}

In this paper we used prescriptions for the ``hydrodynamical'' closure relations that define $\delta P$ and $\pi$ in terms of
other variables, in order to complete the system of equations. Instead of defining the dark energy momentum tensor, it is also
possible to introduce functions that describe the change to the matter growth rate \cite{Amendola:2004wa,Linder:2005in} or that modify the
Einstein equations with an effective Newton's constant and a gravitational slip \cite{Amendola:2007rr}. The latter
parametrisation is in principle equivalent to giving $\delta P$ and $\pi$ as shown explicitly
in \cite{Ballesteros:2011cm}, but the modified growth rate on its own is not sufficient and needs to be supplemented 
by an additional condition. We will call these approaches `modified growth' parameterisations, see also section 1.3.2
of \cite{Amendola:2012ys} for a more detailed introduction.

As the modified growth approach is quite popular and since several groups have derived predictions for the accuracy
with which these parameters can be measured (e.g.\ \cite{Amendola:2007rr,Pogosian:2010tj,Bean:2010zq}), we give here the expressions necessary to compute these
quantities in general and discuss the links between them and our parametrisation. We then show what bounds we
can infer on these modified growth parameters from the data that we use, in the context of our model.

\subsection{Definition of the modified growth parameters}

In general the presence of a dark energy fluid or of a modification of General Relativity will affect the growth rate of the
dark matter perturbations.
We define the growth factor $g$ as the logarithmic derivative of the comoving matter density perturbation,
\be
  g\ \equiv\ \frac{d\log \Delta_m}{d\log a}
\ee
The growth factor is often approximated using the growth index, $\gamma$, as
\be
  g\ =\ \Omega_m(a)^\gamma
\ee
where $\Omega_m(a)\equiv 8\pi Ga^2\rho_m(a)/(3\H^2)$. In general, $g$ and $\gamma$ are space and time dependent functions. To investigate $\gamma$ we express it in terms of $g$
\be
  \gamma\ =\ \frac{\log g}{\log \Omega_m(a)}
\ee
and we have implemented these expressions in CAMB so that we can obtain limits on $g$ and $\gamma$
as derived parameters from our MCMC chains.

However, in general we have two scalar degrees of freedom, related to the possibility to choose independent closure relations for
both $\delta P_{de}$ and $\pi_{de}$ in the fluid picture. To model these two degrees of freedom, 
we introduce a parameter $Q$ which describes either an effective Newton's constant $Q G$ or an additional contribution
to the clustering from $\Delta_{de}$ through the Poisson equation for $\phi$.
In addition, we parameterise the gravitational slip (the difference of the gravitational potentials which is in our model due to the
anisotropic stress of the dark energy) as $\eta$,
\be
Q \equiv \frac{- k^2\phi}{4\pi G a^2 \rho_m \Delta_m}
= 1+ \frac{\rho_{de} \Delta_{de}}{\rho_m \Delta_m}
\,, \qquad
\eta \equiv \frac{\phi}{\psi} \, .
\ee
$\eta$ should not be confused with conformal time, obviously. This parameter is occasionally called $\varpi$ in the literature.

We can now derive some relations between different parameters. For example we have that
\be
\frac{1}{\eta} = 1+ \frac{2}{Q} \frac{\rho_{de} \pi_{de}}{\rho_m \Delta_m} = 1+ \frac{2 \rho_{de} \pi_{de}}{\rho_m \Delta_m + \rho_{de} \Delta_{de}} \, .
\ee
In a pure model 1 situation, i.e.\ with $f_\pi=0$, we then have that
\be
\frac{1}{\eta} = 1+ \frac{2 e_\pi}{Q} \frac{\rho_{de} }{\rho_m } = 1+ \frac{2 e_\pi}{Q} \frac{\Omega_{de}}{\Omega_m} a^{-3 w} \approx 1+ \frac{2 e_\pi}{Q} \, ,
\label{eq:q_eta}
\ee
where the final expression is valid at late times for our averaging (see below). 

\subsection{Constraints on the modified growth parameters in our model}

In general the modified growth parameters are functions of scale and time. We limit here our investigation to the late-time behaviour,
by averaging the parameters over the range $z=0\ldots 1$ using 10 values linearly spaced in $z$. The scale dependence can be
important, so we consider separately `large' scales, $k = 10^{-3} \,h\, \Mpc^{-1}$ and `small' scales, $k = 10^{-1} \,h\, \Mpc^{-1}$. 

\begin{figure}[tb]
\centering
\includegraphics[width=0.32\textwidth]{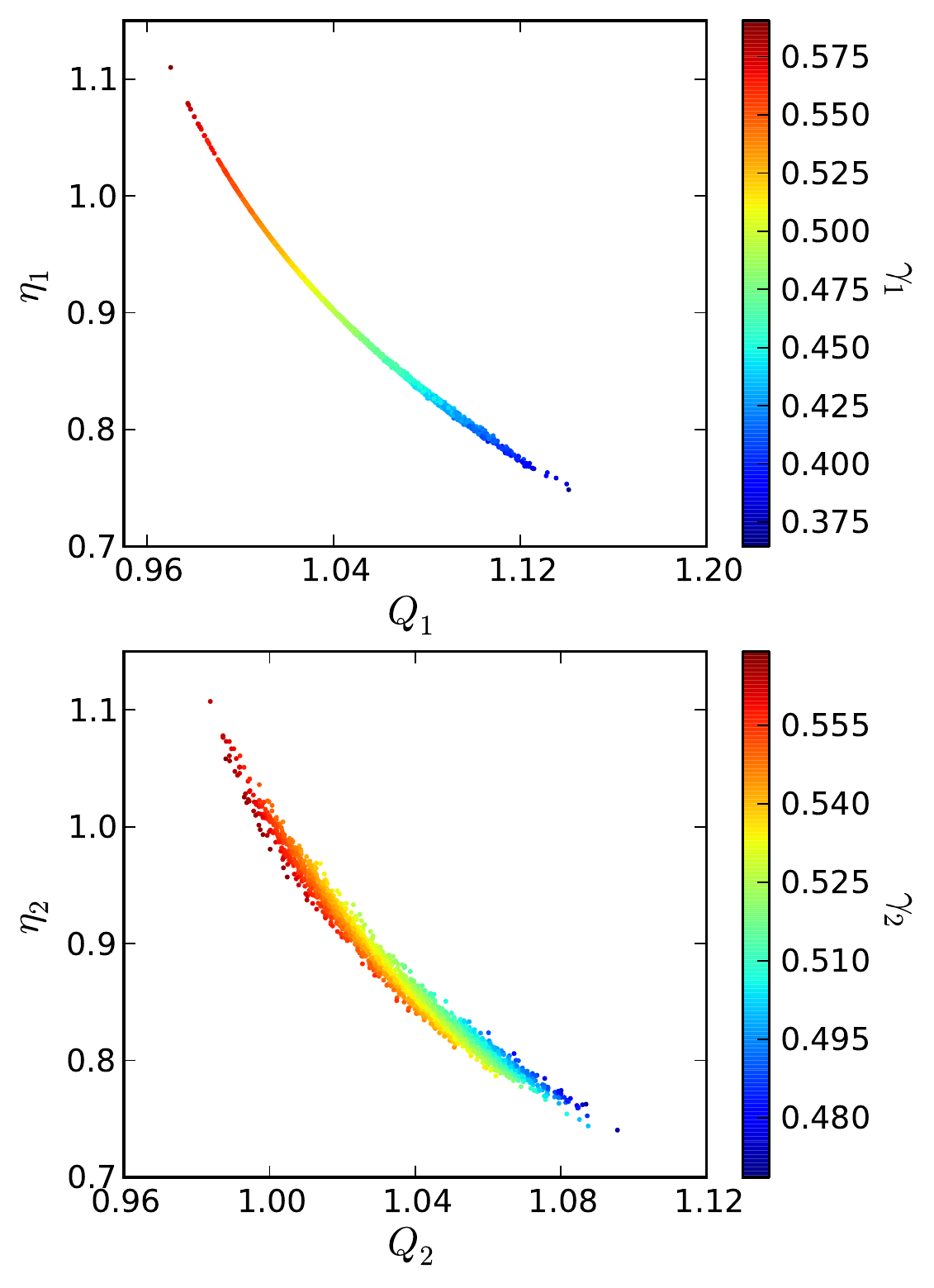}
\includegraphics[width=0.32\textwidth]{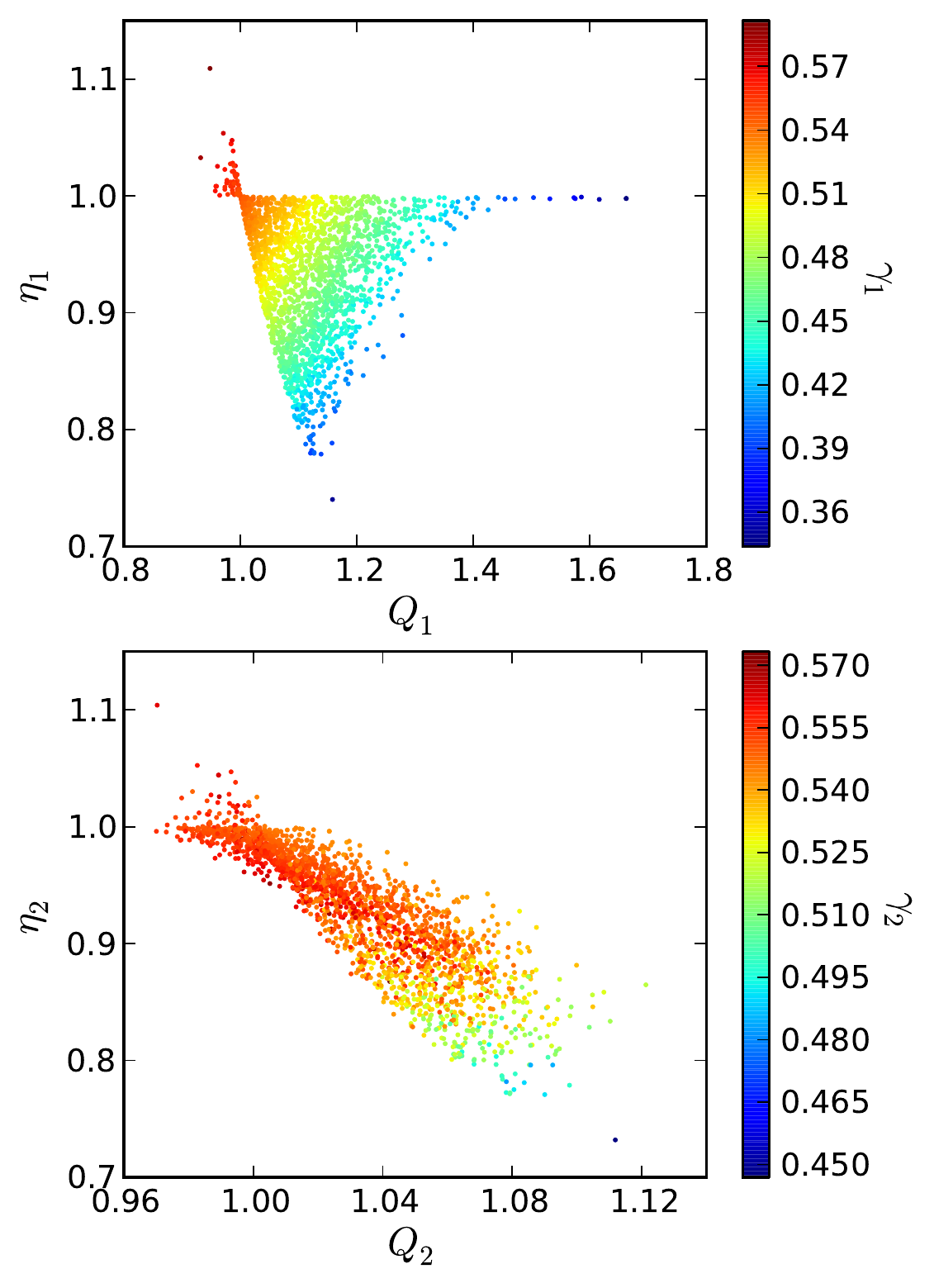}
\includegraphics[width=0.32\textwidth]{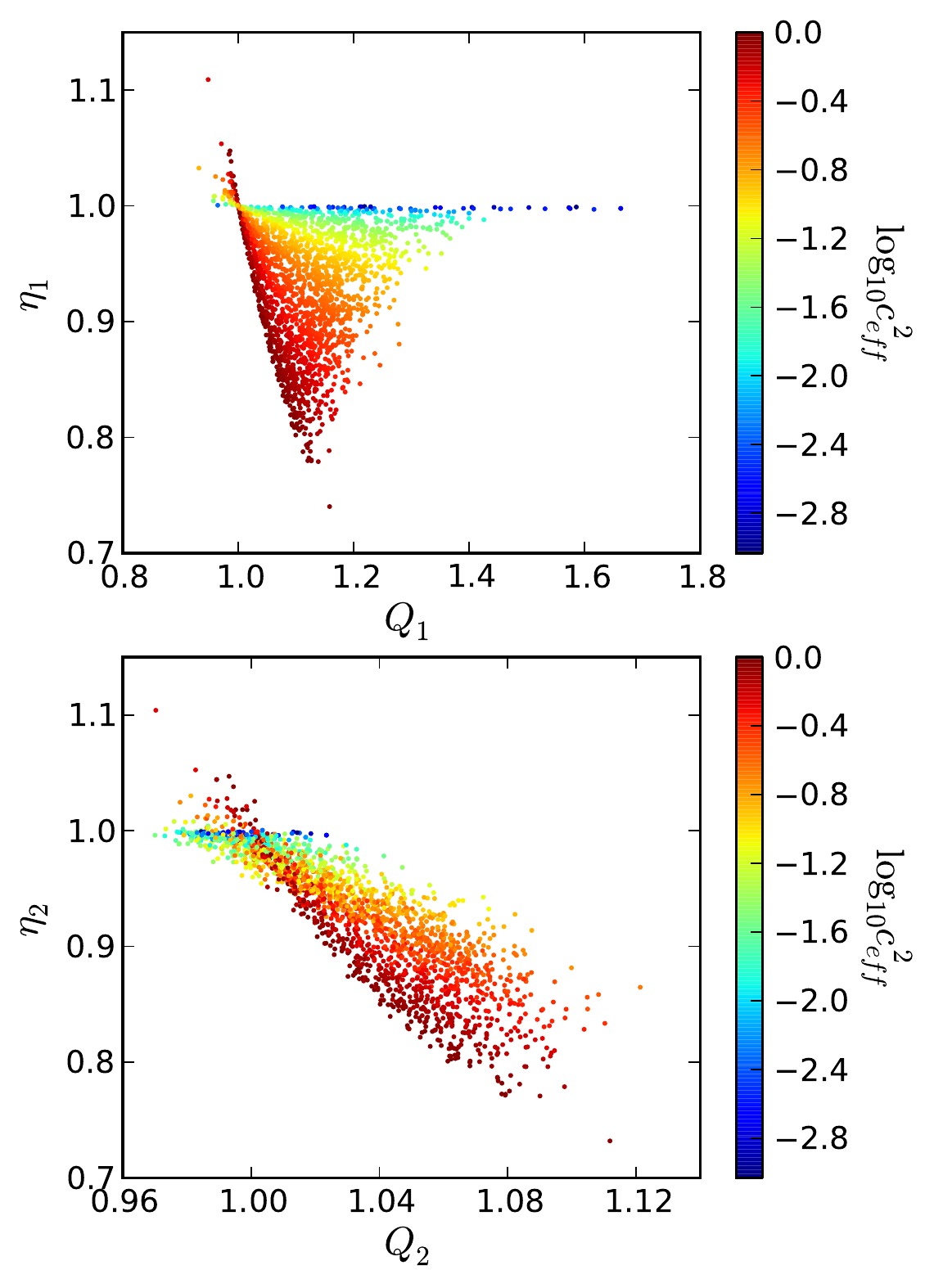}
\caption{Scatter plots of samples of accepted models in our MCMC chains for $\ff=0$ and $\aa$ varying,  when using the full data set, \emph{CMB+BAO+SNe}. The lower row of figures shows the behaviour on large scales $( k_1 = 10^{-3} \,h\,\Mpc^{-1})$, while the upper row depicts smaller scales $ (k_2 = 10^{-1} \,h\,\Mpc^{-1}) $.  \emph{First column:} the sound speed is held fixed, $c_s=1$, and the growth index, $\gamma$, nicely parametrises the single allowed line in the $(Q,\,\eta)$ plane. \emph{Second column:} the sound speed is allowed to vary, a whole area is sampled in the $(Q,\,\eta)$ plane, and the growth index only parametrises one direction. \emph{Third column:} the other direction is parametrised by the sound speed.}
\label{fig:3d}
\end{figure}

In Fig.\ \ref{fig:3d} we show $Q$ and $\eta$ values of a sample of `type 1' models accepted by the MCMC algorithm where $\aa$ varies and $\ff$ is zero. In the first column
we also fix $c_s=1$ (which is equivalent to $\ceff=1$ as $\ff=0$), and in this case we can access only a narrow region in $(Q,\eta)$
space. This is not unexpected as in general we need to vary both $\delta P_{de}$ and $\pi_{de}$. When doing so in the second and third column,
and now a much larger part of the $(Q,\eta)$ parameter space is accessible.  This also illustrates that our models are able to probe
quite generally the space of modifications of the growth parameters.

We can also see very nicely from the colours in the first two columns of Fig.\ \ref{fig:3d} how a non-zero $\aa$ changes the growth
rate $\gamma$, with pretty much a one-to-one mapping between the two on small scales. The sound speed on the other hand leads to a rotation
in the $(Q,\eta)$ parameter space on small scales.

We can understand this latter behaviour with the help of the relation (\ref{eq:q_eta}): The presence of an anisotropic stress induced
by $\aa$ impacts not only $\eta$, but also $Q$. On sub-sound-horizon scales and during matter domination the induced dark energy
perturbations are given by Eq.\ (\ref{eq:pheno12}),
\be
\Delta_{de} \approx \delta_{de} \approx a \delta_0  \left( \frac{2 e_\pi a^n}{3 \ceff^2} \right) \approx \left( \frac{2 e_\pi a^n}{3 \ceff^2} \right) \Delta_m \, ,
\ee
and therefore we have in that limit that
\be
Q = 1 + \frac{2 \aa a^n}{3 \ceff^2} \frac{\rho_{de}}{\rho_m} \, \Rightarrow \, \frac{1}{\eta} = 1+ \frac{6 \ceff^2 \aa a^n \rho_{de}}{2 \aa a^n \rho_{de} + 3 \ceff^2 \rho_m} \, .
\label{eq:q_eta_2}
\ee
For high sound speed, $c_s\approx 1$, we find from the MCMC exploration that at 95\% CL $-0.01 < \aa < 0.18$, and the allowed region shrinks around $\aa=0$ as the sound
speed decreases, see the red contours in Fig.\ \ref{fig:2d_models}. Plotting 
the curves $Q(\aa;\ceff)$ and $\eta(\aa;\ceff)$ for the allowed values of $\aa$ leads to a figure that corresponds very well to the region visible in the upper row of Fig.\ \ref{fig:3d}.
We notice that for $\aa=0$ we have that $Q=1$ and $\eta=1$ independently of the sound speed. This behaviour is clearly
visible in the figure. We can also see that as $\ceff \rightarrow 0$ the slip vanishes, $\eta\rightarrow 1$, even if $\aa\neq0$,
while the impact of $\aa$ on $Q$ is enhanced, explaining the horizontal line visible in the top right-hand panel of Fig.\ \ref{fig:3d}
for low sound speed.

On large scales, the impact of $\ceff$ is more indirect, by changing the size of the sound horizon. We can see that for a lower effective
sound speed we have a larger $Q$ for a given $\eta$, as the dark energy is able to cluster more easily. 
On super-sound (but sub-horizon) scales and during matter dominance, dark energy density perturbations are given by (Eq. (\ref{eq:pheno9}))
\be 
\Delta_{de}  &\approx &  \frac{2 \aa a^n \Delta_m}{3 \[ 2\( 1+\alpha \) + \vartheta + n(3+\alpha+n) \]} \frac{k^2}{\H^2} \\
\label{eq:parametrisations:1}
Q &=& 1 + \frac{2 \aa a^n}{3}\frac{\rho_{de}}{\rho_m} \frac{k^2}{\H^2} \, \\
\frac{1}{\eta} &=& 1 + \frac{6 \aa a^n \rho_{de} }{3 \rho_m  + 2 \aa a^n \rho_{de} \frac{k^2}{\H^2} \[ 2\( 1+\alpha \) + \vartheta + n(3+\alpha + n)\]^{-1}}\, .
\label{eq:parametrisations:2}
\ee

On the other hand, if we consider solutions on super-horizon scales, that is, Eq. (\ref{eq:pheno21})-(\ref{eq:pheno23}), we have 
\be 
\Delta_{de} \approx \Delta_m \[  \frac{4 \aa a^n - 3(1+w)}{-3} \] \, ,
\label{eq:parametrisations:3}
\ee 
\be 
Q =  1 + \[ -\frac{4}{3} \aa a^n + (1+w)\] \frac{\rho_{de}}{\rho_m} \quad \Rightarrow \quad \frac{1}{\eta} = 1 + \frac{6 \aa a^n \rho_{de}}{3 \rho_m - \[ 4\aa a^n - 3(1+w)\]\rho_{de}}
\label{eq:parametrisations:4}
\ee
The last equation shows that on horizon scales dark energy can cluster even if $\aa=0$. This is the reason why 
there is no `clean' intersection of the curves at $Q=1$, $\eta=1$ in the lower row of Fig.\ \ref{fig:3d}.


\section{Conclusions}

In this paper we study effective fluid dark energy models that have a non-zero anisotropic stress $\pi_{de}$. These models
can represent not only dark energy, but also modified gravity models \cite{Kunz:2006ca}. We consider specifically two
scenarios, one where the dark energy anisotropic stress is linked to the dark matter density perturbations by a parameter $\aa$, and
another model where $\pi_{de}$ is linked to the dark energy density perturbations by a parameter $\ff$. These are only two out
of a range of possibilities that arise naturally in general models like the Horndeski Lagrangian, but we think that
they illustrate rather well the impact of a non-zero anisotropic stress that is either internal to the dark energy (model 2)
or externally sourced (model 1). In addition we allow for a free sound speed $c_s$ for the dark energy perturbations.

When studying the evolution of the perturbations, we find that
the internal anisotropic stress changes the effective sound speed of the dark energy, see Eq.\ (\ref{eq:pheno2}).
This means that the anisotropic stress can stabilise the dark energy perturbations even if $c_s^2<0$, but also that
$c_s^2>0$ does not guarantee stability, as the relevant quantity is $\ceff^2=c_s^2 - 2\ff / 3$. We also find that a sufficiently negative
$\pi_{de}$ (relative to $\Delta_{de}$) can lead to rapid growth of the dark energy perturbations in the regime that is sub-horizon
but outside of the sound horizon (cf Tab.\ \ref{tab:stability}).

We further find that the contribution to $\pi_{de}$ from $\Delta_m$ acts like
an external source of dark energy perturbations. This coupling can lead to growing perturbations both inside the
dark energy sound horizon and outside of the Hubble horizon, at least as long as the dark matter is dominating the
evolution of the universe. With the purely `internal' anisotropic stress of our
model 2 (where $\pi_{de} \sim \Delta_{de}$) this does not happen. If the coupling to the matter perturbations is zero,
then the dark energy perturbations become constant on sub-sound horizon or super-Hubble horizon scales during
matter domination, even in the presence of an internal $\pi_{de}$ (except when the effective sound speed of the
dark energy becomes imaginary).

For all of these special cases we provide analytical approximations for the behaviour of the dark energy perturbations.
On the one hand, these are useful to understand the behaviour of the dark energy and the resulting observational constraints, and on the other hand, they can be used to correctly set initial conditions for numerical codes. 

When looking at the constraints from the cosmic microwave background, augmented by distance data from BAO and SN-Ia,
we find that the external contribution to $\pi_{de}$ is quite well constrained, $-0.01 < \aa < 0.13$ at 95\% CL for $\ff=0$ (marginalised 
over $\log(\ceff)$), and $-0.01 < \aa < 0.23$ when also marginalising over $\ff$ and $\gg$,
see Fig.\ \ref{fig:2d_models}. The internal contribution 
is much less constrained, and is limited mostly by the stability of the perturbations. We also considered the resulting
constraints on the `modified growth' parameters like the growth index $\gamma$, the effective Newton's constant $Q$
and the gravitational slip $\eta$, shown in Fig.\ \ref{fig:3d}. 

Overall, adding anisotropic stress to dark energy models (effectively turning them into modified gravity models \cite{Saltas:2010tt})
opens up a new region of parameter space that is poorly constrained by the primary CMB anisotropies alone. Constraining these
models requires additional data that probes the evolution of the perturbations, like weak lensing observations, redshift space
distortions, the galaxy distribution and the growth rate of structure. Currently ongoing and future experiments will provide a wealth
of data to improve our understanding of the dark energy, but it is important that the data sets are analysed carefully and consistently
by taking into account the full cosmological model, without assuming $\Lambda$CDM or smooth dark energy from the beginning.

\acknowledgments
We would like to thank Ruth Durrer and Antonio Riotto for helpful discussions and Adam Riess for useful comments. 
W.C.\ is supported by the Administrative Department of Science, Technology and Innovation COLCIENCIAS. 
M.K.\ is supported by the Swiss National Science Foundation. 
L.H.\ acknowledges financial support by the Norwegian Research Council and the Swiss National Science Foundation in the early stages of this work. 
Part of the numerical calculations for this project were run on the Andromeda cluster of the University of Geneva.

The development of Planck has been supported by: ESA; CNES and CNRS/INSU-IN2P3-INP (France); ASI, CNR, and INAF (Italy); NASA and DoE (USA); STFC and UKSA (UK); CSIC, MICINN and JA (Spain); Tekes, AoF and CSC (Finland); DLR and MPG (Germany); CSA (Canada); DTU Space (Denmark); SER/SSO (Switzerland); RCN (Norway); SFI (Ireland); FCT/MCTES (Portugal); and The development of Planck has been supported by: ESA; CNES and CNRS/INSU-IN2P3-INP (France); ASI, CNR, and INAF (Italy); NASA and DoE (USA); STFC and UKSA (UK); CSIC, MICINN and JA (Spain); Tekes, AoF and CSC (Finland); DLR and MPG (Germany); CSA (Canada); DTU Space (Denmark); SER/SSO (Switzerland); RCN (Norway); SFI (Ireland); FCT/MCTES (Portugal); and PRACE (EU).

A description of the Planck Collaboration and a list of its members, including the technical or scientific activities in which they have been involved, can be found at\\ \href{http://www.sciops.esa.int/index.php?project=planck&page=Planck_Collaboration}{http://www.sciops.esa.int/index.php?project=planck\&page=Planck\_Collaboration}.


\appendix

\section{Stability}

The non-autonomous system of coupled differential equations (\ref{eq:de-cont})-(\ref{eq:de-eul}) for density and velocity perturbations can be written as
\be 
\vec{\mathbf{x}}' = \mathbf{B}(a;\theta_j) \, \vec{\mathbf{x}}\, ,
\label{eq:appendix:B1}
\ee
where $ \vec{\mathbf{x}}^\intercal = (\delta_m,\, V_m,\, \delta_{de},\, V_{de}) $, $ \mathbf{B} $ is a $ 4\times4 $ matrix which depends on the parameters $\theta_j=\lcb H_0,\, w,\, \Omega_m,\, \Omega_x,\,  c_s,\, \aa,\, \ff,\, \gg \rcb$, on the mode $ k $ and the scale factor $ a $. In order to obtain information about the parameter space, we assess the stability of the system (\ref{eq:appendix:B1}). We compute the eigenvalues $ \lambda_k (a,\, \theta_j)$ of the matrix $\mathbf{B}(a;\theta_j)$ numerically and look at the regions in the $\theta_j-$space where all eigenvalues have negative real parts for the whole time interval we consider, that is, from matter domination to the present time. 

\begin{figure}[tb]
\centering
\includegraphics[scale=1]{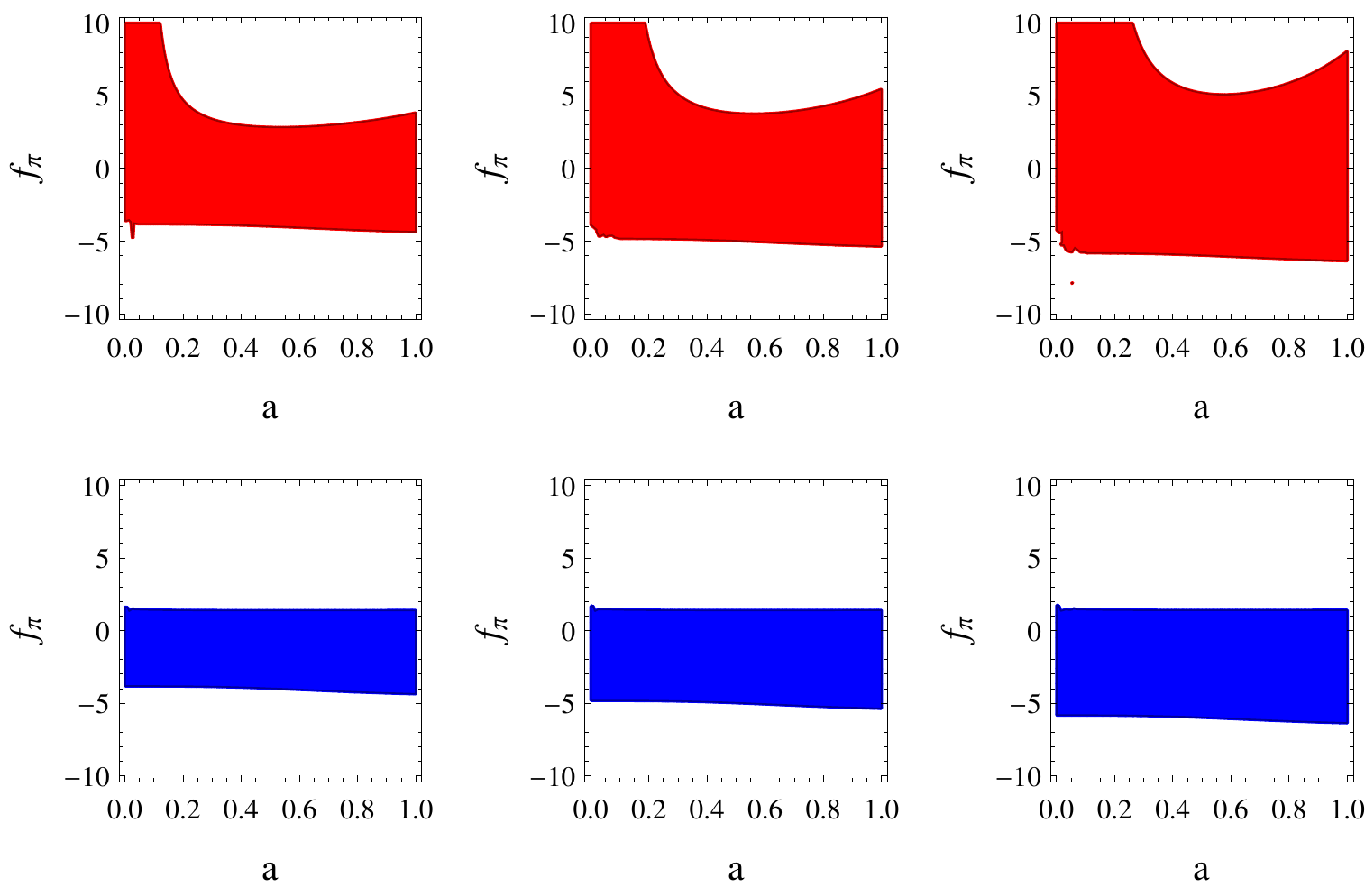} 
\caption{The figure shows the region in the $ \ff-$space for which all the eigenvalues of the matrix $ \mathbf{A} $ in Eq. (\ref{eq:appendix:B2}) have negative real parts. The red region (upper panel) corresponds to a mode on super-horizon scales ($k=5 H_0$) in matter domination. The blue region (lower panel) shows a sub-horizon mode  ($ k=300 H_0 $). We have used $c_s^2 = 1$, $w=-1.05$, $\aa=0$ and $\gg=0$. In each column, from left to right, we rescale the variables by using powers $ m=2,\,3,\,4 $.}
\label{fig:stability}
\end{figure}

For the system (\ref{eq:appendix:B1}) we find one eigenvalue which does not have a region in $\theta_j-$space where $\operatorname{Re} \left( \lambda_k (a,\, \theta_j) \right) < 0$. Then, since we know that in matter dominated era and on sub-horizon scales matter perturbations grow linearly, that is, $ \delta_m \propto a$, we use the following approach to assess the stability of the system. First, we rescale the  variables $ \vec{\mathbf{x}} $ dividing them by a power $ a^m $ of the scale factor, $\vec{\mathbf{y}}=a^{-m}\vec{\mathbf{x}}$. It follows that the system  (\ref{eq:appendix:B1}) becomes   
\be
\vec{\mathbf{y}}' = \mathbf{A}(a;\theta_j) \, \vec{\mathbf{y}}\, ,
\label{eq:appendix:B2}
\ee
where $ \mathbf{A}(a;\theta_j) = \mathbf{B}(a;\theta_j) - \frac{m}{a} \mathbf{I}  $, with $ \mathbf{I} $ the identity matrix. Second, we find regions in parameter space where all the eigenvalues of the matrix $ \mathbf{A} $ have negative real parts. We study models with $ \aa=\gg=0 $ since the effective sound speed $ \ceff^2 $ is only defined in terms of $ c_s $ and $ \ff $;  moreover we check the robustness of our method with different powers $ m $. Fig. \ref{fig:stability} shows regions where all eigenvalues have real part negative in the time interval relevant for both super and sub-horizon scales.  Since we use $ c_s^2=1 $ we can see clearly an upper limit on $ \ff =3/2$, which nicely agrees with $ \ceff^2>0 $.  

\begin{figure}[tb]
\centering 
\includegraphics[scale=1]{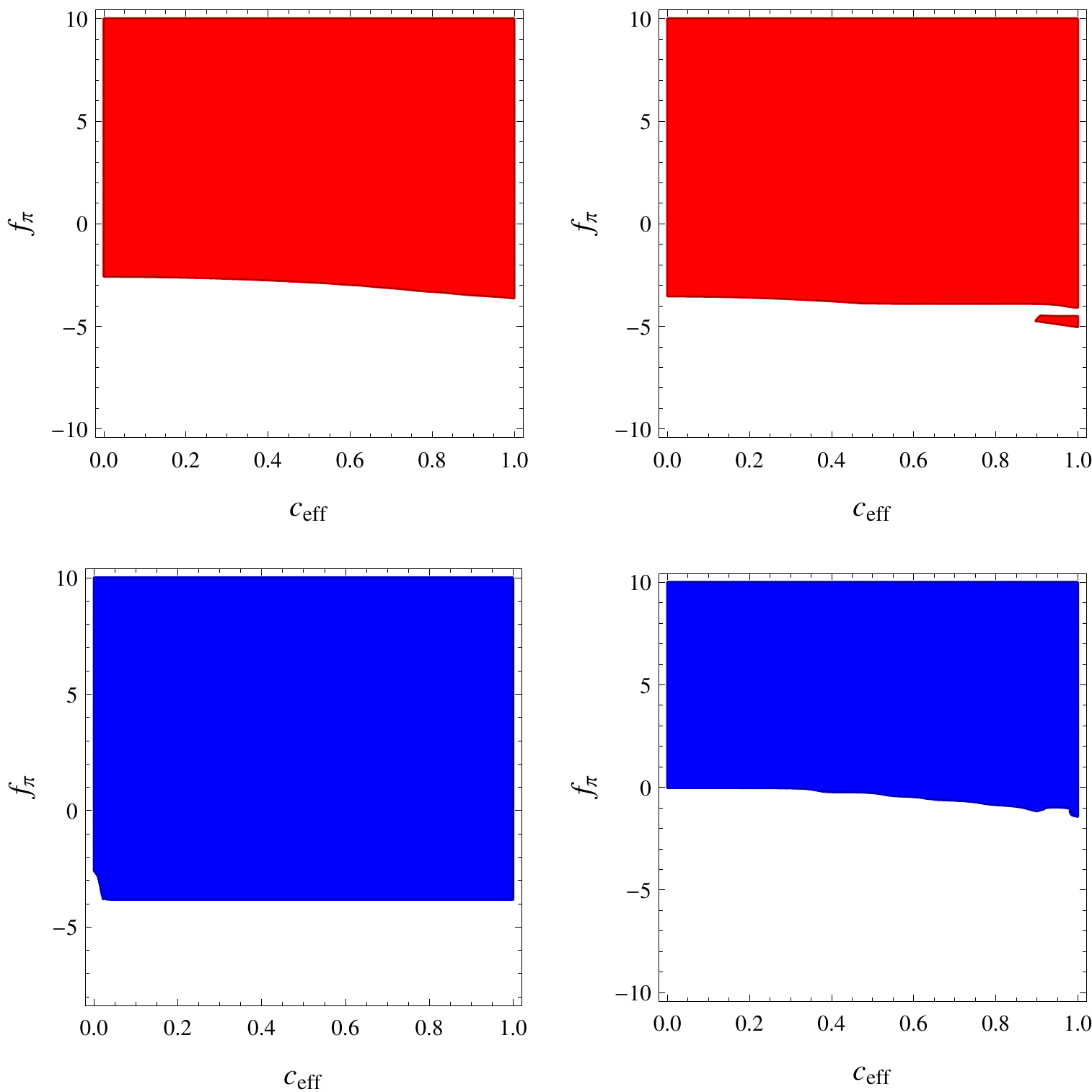} 
\caption{Stability regions: parameter regions where the perturbations grow more slowly than $a^2$ at $a=5\times10^{-2}$, for $w=-1.05$ and $\aa=0$. The upper row (red contours) are for $k=5 H_0$, while the lower row (blue contours) is for $k=300 H_0$. The right panels are for $ \gg = 10^{-5} $ and the left panels for $ \gg=10^5 $.}
\label{fig:stability:2}
\end{figure}

Then we study the impact of the parameter $ \gg $ on the stability of the system. For a given scale factor $ a $ we determine regions in the plane $ \ceff-\ff $ for which the system (\ref{eq:appendix:B2}) is stable for both $ \gg \mug 1 $ and $ \gg \ll 1 $. Fig. \ref{fig:stability:2} shows again the stable regions for large and small scales. 

\section{General solutions}\label{appendix:2}

In Sec.\ \ref{section:3} we study some limiting cases in the $ 4- $dimensional system (\ref{eq:de-cont})-(\ref{eq:de-eul}) for which dark matter and dark energy perturbations decouple from each other.  In this appendix we give some solutions which are a bit cumbersome to be written in the main body of the paper. 

On sub-horizon scales and during matter dominance, dark energy density perturbations are governed by Eq. (\ref{eq:pheno7}) whose full solution is

\bea
\label{eq:appendix:A1}
\delta_{de} & = & \(\frac{\ceff k}{\H}\)  \, a^{\alpha_8}\,\lcb A_5\,  J_{-\nu_5}\( \frac{2 \ceff k}{\H} \) + A_6\, J_{\nu_5}\( \frac{2 \ceff k}{\H} \) \rcb  \nonumber \\  
&+& \aa \beta_6\, a^{\alpha_7} \( \frac{\ceff k}{\H} \)^{\alpha_6} \,  J_{\nu_5}\( \frac{2 \ceff k}{\H} \) \, \tilde{{}_1F_2} \( \nu_6\,;\, \alpha_6\,, \nu_7\, ; -\frac{k^2 \ceff^2}{\H^2} \)   \nonumber\\ 
& - & \aa \beta_7 a^{\alpha_7} \( \frac{\ceff k}{\H} \)^{\alpha_{9}} J_{-\nu_5}\(  \frac{2\ceff k}{\H} \) \tilde{{}_1F_2} \( \nu_8\,;\, \alpha_{9}\,, \nu_{9}\, ; -\frac{k^2 \ceff^2}{\H^2} \)  \,,  
 \eea

\noindent where 

\bea 
\nu_5 & = & \frac{\sqrt{432\ceff^4 + 48\ff^2 + 72\ceff^2(-1+4\ff-6w) + 3(1+6w)^2 -8\ff(3+4\gg^2+ 18w)}}{2\sqrt{3}}\,,\\
\nu_6 & = & -\frac{5}{4} -\ff + n + \frac{3w-\nu_5}{2}\,,
\eea

\bea
\nu_7 & = & \nu_6 + 1\,,\\
\nu_8 & = &  -\frac{5}{4} -\ff + n + \frac{3w+\nu_5}{2}\,,\\
\nu_9 & = & \nu_8 + 1\,,\\
\alpha_6 & = & 1 - \nu_5\,, \\
\alpha_7 & = & -2(1 +\ff) +3w +n\,,\\
\alpha_8 & = &  -\ff + \frac{3w}{2} - \frac{3}{4}\,, \\
\alpha_{9} & = & 1 + \nu_5\,,
\eea

\bea
\beta_6 & = & \frac{1}{24} \frac{\ceff^{-2-4\ff+6w} k^{-4\ff+6w} }{H_0^{2-4\ff+6w}}   \Omega_m^{-1+2\ff-3w} \pi \delta_0 \csc \( \pi \nu_5\) \Gamma \( \alpha_6\) \Gamma \( \nu_6\) \Gamma \( \alpha_{9}\)\,,\\
\beta_7 & = & \frac{1}{24} \frac{\ceff^{-2-4\ff+6w} k^{-4\ff+6w} }{H_0^{2-4\ff+6w}}   \Omega_m^{-1+2\ff-3w} \pi \delta_0 \csc \( \pi \nu_5\) \Gamma \( \alpha_6\) \Gamma \( \nu_9\) \Gamma \( \alpha_{9}\)\,,
\eea

\noindent $ A_5$, $ A_6  $ are constants of integration and $ \tilde{{}_1F_2} $ stands for the regularized generalized hypergeometric function. The last two terms in Eq. (\ref{eq:appendix:A1}) are due to the external anisotropic stress.

On the other hand, for sub-sound horizon scales, during dark energy domination and without external contribution to the dark energy anisotropic stress ($\aa=0$), we find (by using Eq. (\ref{eq:pheno19})) for dark energy velocity perturbations
\begin{align}
\label{eq:appendix:A2}
V_{de} = & \frac{1}{2} \( \frac{x_3}{2} \)^{\frac{1-\alpha_4}{1+3w}} \Bigg\{\,
  \bigg[\, 6(c_s^2-w) + 1-\alpha_4 \,\bigg] \times
 \nonumber \\ 
 &\times \[ B_4 \, \Gamma\(\frac{2+3w-\alpha_4}{1+3w}\)   J_{-\nu_1}(x_3)   + A_4\, \Gamma\(\frac{3w+\alpha_4}{1+3w}\)  J_{\nu_1}(x_3) \] \nonumber \\
&+ \frac{x_3 (1+3w)}{4} \[ B_4\,\Gamma\(\frac{2+3w-\alpha_4}{1+3w}\) \[ J_{-1-\nu_1}(x_3) - J_{1-\nu_1}(x_3)  \]    \right.
\nonumber \\
&+ \left. A_4\, \Gamma\(\frac{3w+\alpha_4}{1+3w}\) \[ J_{-1+\nu_1}(x_3) - J_{1+\nu_1}(x_3) \]    \]       \,\Bigg\} \, , 
\end{align}
while for dark matter velocity perturbations 
\bea 
\label{eq:appendix:A3}
V_m & = & \frac{3 (1+2\ff)}{1+3w} 2^{-\frac{2-\alpha_4 + \nu_1 +3w(1+\nu_1)}{1+3w}} x_3^{-\frac{-1+\alpha_4+\nu_1+3w\nu_1}{1+3w}} \,\Bigg\{
A_4 \, x_3^{2\nu_1}
 \\
& \times & \Gamma\(\frac{3w+\alpha_4}{1+3w}\) \Gamma\( \frac{2-\alpha_4 +3w(\nu_1 -1)+\nu_1}{2+6w} \)
\nonumber \\ 
&\times & {}_1F_2\( \lcb \frac{2-\alpha_4 + 3w(\nu_1-1) + \nu_1}{2+6w} \rcb,\, \lcb 1+\nu_1,\, \frac{4-\alpha_4+\nu_1+3w(1+\nu_1)}{2+6w}  \rcb,\, -\frac{1}{4}x_3^2  \)
 \nonumber \\
&+& B_4\, 4^{\nu_1}\, \Gamma\( \frac{2+3w-\alpha_4}{1+3w} \) \Gamma\(-\frac{-2+\alpha_4 + \nu_1 +3w(1+\nu_1)}{2+6w} \)
\nonumber \\
&\times & {}_1F_2\( \lcb \frac{2-\alpha_4 - 3w(\nu_1+1) - \nu_1}{2+6w} \rcb,\, \lcb 1-\nu_1,\, \frac{4-\alpha_4-\nu_1-3w(-1+\nu_1)}{2+6w}  \rcb,\, -\frac{1}{4}x_3^2  \)  \,\Bigg\}
\nonumber \, ,
\eea
where $ \alpha_4,\, \nu_1  $ and $ x_3 $ are given by Eq. (\ref{eq:pheno20}). By means of this solution we can easily find an expression for dark matter density perturbations by solving the differential equation 
\be
\delta_m' = \frac{V_m}{a} \,.
\label{eq:appendix:A4}
\ee


\bibliography{ade_refs}

\providecommand{\href}[2]{#2}\begingroup\raggedright\begin{thebibliography}{10}

\bibitem{Copeland:2006wr}
E.~J. Copeland, M.~Sami, and S.~Tsujikawa, {\it {Dynamics of dark energy}},
  {\em Int.J.Mod.Phys.} {\bf D15} (2006) 1753--1936,
  [\href{http://xxx.lanl.gov/abs/hep-th/0603057}{{\tt hep-th/0603057}}].

\bibitem{Durrer:2008in}
R.~Durrer and R.~Maartens, {\it {Dark Energy and Modified Gravity}},
  \href{http://xxx.lanl.gov/abs/0811.4132}{{\tt arXiv:0811.4132}}.

\bibitem{Frieman:2008sn}
J.~Frieman, M.~Turner, and D.~Huterer, {\it {Dark Energy and the Accelerating
  Universe}},  {\em Ann.Rev.Astron.Astrophys.} {\bf 46} (2008) 385--432,
  [\href{http://xxx.lanl.gov/abs/0803.0982}{{\tt arXiv:0803.0982}}].

\bibitem{Amendola2010}
L.~Amendola and S.~Tsujikawa, {\em {Dark Energy: Theory and Observations}}.
\newblock Cambridge University Press, 2010.

\bibitem{Clifton:2011jh}
T.~Clifton, P.~G. Ferreira, A.~Padilla, and C.~Skordis, {\it {Modified Gravity
  and Cosmology}},  {\em Phys.Rept.} {\bf 513} (2012) 1--189,
  [\href{http://xxx.lanl.gov/abs/1106.2476}{{\tt arXiv:1106.2476}}].

\bibitem{Amendola:2012ys}
{\bf Euclid Theory Working Group} Collaboration, L.~Amendola et~al., {\it
  {Cosmology and fundamental physics with the Euclid satellite}},  {\em Living
  Rev.Rel.} {\bf 16} (2013) 6, [\href{http://xxx.lanl.gov/abs/1206.1225}{{\tt
  arXiv:1206.1225}}].

\bibitem{Kunz:2012aw}
M.~Kunz, {\it {The phenomenological approach to modeling the dark energy}},
  {\em Comptes Rendus Physique} {\bf 13} (2012) 539--565,
  [\href{http://xxx.lanl.gov/abs/1204.5482}{{\tt arXiv:1204.5482}}].

\bibitem{Battye:2012eu}
R.~A. Battye and J.~A. Pearson, {\it {Effective action approach to cosmological
  perturbations in dark energy and modified gravity}},  {\em JCAP} {\bf 1207}
  (2012) 019, [\href{http://xxx.lanl.gov/abs/1203.0398}{{\tt
  arXiv:1203.0398}}].

\bibitem{Sawicki:2012re}
I.~Sawicki, I.~D. Saltas, L.~Amendola, and M.~Kunz, {\it {Consistent
  perturbations in an imperfect fluid}},  {\em JCAP} {\bf 1301} (2013) 004,
  [\href{http://xxx.lanl.gov/abs/1208.4855}{{\tt arXiv:1208.4855}}].

\bibitem{Baker:2012zs}
T.~Baker, P.~G. Ferreira, and C.~Skordis, {\it {The Parameterized
  Post-Friedmann Framework for Theories of Modified Gravity: Concepts,
  Formalism and Examples}},  {\em Phys.Rev.} {\bf D87} (2013) 024015,
  [\href{http://xxx.lanl.gov/abs/1209.2117}{{\tt arXiv:1209.2117}}].

\bibitem{Gubitosi:2012hu}
G.~Gubitosi, F.~Piazza, and F.~Vernizzi, {\it {The Effective Field Theory of
  Dark Energy}},  {\em JCAP} {\bf 1302} (2013) 032,
  [\href{http://xxx.lanl.gov/abs/1210.0201}{{\tt arXiv:1210.0201}}].

\bibitem{Bloomfield:2012ff}
J.~K. Bloomfield, E.~E. Flanagan, M.~Park, and S.~Watson, {\it {Dark energy or
  modified gravity? An effective field theory approach}},  {\em JCAP} {\bf
  1308} (2013) 010, [\href{http://xxx.lanl.gov/abs/1211.7054}{{\tt
  arXiv:1211.7054}}].

\bibitem{Amendola:2007rr}
L.~Amendola, M.~Kunz, and D.~Sapone, {\it {Measuring the dark side (with weak
  lensing)}},  {\em JCAP} {\bf 0804} (2008) 013,
  [\href{http://xxx.lanl.gov/abs/0704.2421}{{\tt arXiv:0704.2421}}].

\bibitem{Hu:2007pj}
W.~Hu and I.~Sawicki, {\it {A Parameterized Post-Friedmann Framework for
  Modified Gravity}},  {\em Phys.Rev.} {\bf D76} (2007) 104043,
  [\href{http://xxx.lanl.gov/abs/0708.1190}{{\tt arXiv:0708.1190}}].

\bibitem{Amendola:2012ky}
L.~Amendola, M.~Kunz, M.~Motta, I.~D. Saltas, and I.~Sawicki, {\it {Observables
  and unobservables in dark energy cosmologies}},  {\em Phys.Rev.} {\bf D87}
  (2013) 023501, [\href{http://xxx.lanl.gov/abs/1210.0439}{{\tt
  arXiv:1210.0439}}].

\bibitem{Motta:2013cwa}
M.~Motta, I.~Sawicki, I.~D. Saltas, L.~Amendola, and M.~Kunz, {\it {Probing
  Dark Energy through Scale Dependence}},  {\em Phys.Rev.} {\bf D88} (2013)
  124035, [\href{http://xxx.lanl.gov/abs/1305.0008}{{\tt arXiv:1305.0008}}].

\bibitem{Kunz:2006ca}
M.~Kunz and D.~Sapone, {\it {Dark Energy versus Modified Gravity}},  {\em
  Phys.Rev.Lett.} {\bf 98} (2007) 121301,
  [\href{http://xxx.lanl.gov/abs/astro-ph/0612452}{{\tt astro-ph/0612452}}].

\bibitem{Saltas:2010tt}
I.~D. Saltas and M.~Kunz, {\it {Anisotropic stress and stability in modified
  gravity models}},  {\em Phys.Rev.} {\bf D83} (2011) 064042,
  [\href{http://xxx.lanl.gov/abs/1012.3171}{{\tt arXiv:1012.3171}}].

\bibitem{Sapone:2010iz}
D.~Sapone, {\it {Dark Energy in Practice}},  {\em Int.J.Mod.Phys.} {\bf A25}
  (2010) 5253--5331, [\href{http://xxx.lanl.gov/abs/1006.5694}{{\tt
  arXiv:1006.5694}}].

\bibitem{Bardeen:1980kt}
J.~M. Bardeen, {\it {Gauge Invariant Cosmological Perturbations}},  {\em
  Phys.Rev.} {\bf D22} (1980) 1882--1905.

\bibitem{Kodama:1985bj}
H.~Kodama and M.~Sasaki, {\it {Cosmological Perturbation Theory}},  {\em
  Prog.Theor.Phys.Suppl.} {\bf 78} (1984) 1--166.

\bibitem{Ma:1995ey}
C.-P. Ma and E.~Bertschinger, {\it {Cosmological perturbation theory in the
  synchronous and conformal Newtonian gauges}},  {\em Astrophys.J.} {\bf 455}
  (1995) 7--25, [\href{http://xxx.lanl.gov/abs/astro-ph/9506072}{{\tt
  astro-ph/9506072}}].

\bibitem{Ballesteros:2011cm}
G.~Ballesteros, L.~Hollenstein, R.~K. Jain, and M.~Kunz, {\it {Nonlinear
  cosmological consistency relations and effective matter stresses}},  {\em
  JCAP} {\bf 1205} (2012) 038, [\href{http://xxx.lanl.gov/abs/1112.4837}{{\tt
  arXiv:1112.4837}}].

\bibitem{Koyama:2005kd}
K.~Koyama and R.~Maartens, {\it {Structure formation in the dgp cosmological
  model}},  {\em JCAP} {\bf 0601} (2006) 016,
  [\href{http://xxx.lanl.gov/abs/astro-ph/0511634}{{\tt astro-ph/0511634}}].

\bibitem{Song:2010rm}
Y.-S. Song, L.~Hollenstein, G.~Caldera-Cabral, and K.~Koyama, {\it {Theoretical
  Priors On Modified Growth Parametrisations}},  {\em JCAP} {\bf 1004} (2010)
  018, [\href{http://xxx.lanl.gov/abs/1001.0969}{{\tt arXiv:1001.0969}}].

\bibitem{Koivisto:2005mm}
T.~Koivisto and D.~F. Mota, {\it {Dark energy anisotropic stress and large
  scale structure formation}},  {\em Phys.Rev.} {\bf D73} (2006) 083502,
  [\href{http://xxx.lanl.gov/abs/astro-ph/0512135}{{\tt astro-ph/0512135}}].

\bibitem{Mota:2007sz}
D.~Mota, J.~Kristiansen, T.~Koivisto, and N.~Groeneboom, {\it {Constraining
  Dark Energy Anisotropic Stress}},  {\em Mon.Not.Roy.Astron.Soc.} {\bf 382}
  (2007) 793--800, [\href{http://xxx.lanl.gov/abs/0708.0830}{{\tt
  arXiv:0708.0830}}].

\bibitem{Sapone:2012nh}
D.~Sapone and E.~Majerotto, {\it {Fingerprinting Dark Energy III: distinctive
  marks of viscosity}},  {\em Phys.Rev.} {\bf D85} (2012) 123529,
  [\href{http://xxx.lanl.gov/abs/1203.2157}{{\tt arXiv:1203.2157}}].

\bibitem{Sapone:2013wda}
D.~Sapone, E.~Majerotto, M.~Kunz, and B.~Garilli, {\it {Can dark energy
  viscosity be detected with the Euclid survey?}},  {\em Phys.Rev.} {\bf D88}
  (2013) 043503, [\href{http://xxx.lanl.gov/abs/1305.1942}{{\tt
  arXiv:1305.1942}}].

\bibitem{Hu:1998kj}
W.~Hu, {\it {Structure formation with generalized dark matter}},  {\em
  Astrophys.J.} {\bf 506} (1998) 485--494,
  [\href{http://xxx.lanl.gov/abs/astro-ph/9801234}{{\tt astro-ph/9801234}}].

\bibitem{Sapone:2009kx}
D.~Sapone and M.~Kunz, {\it Fingerprinting dark energy},  {\em Phys. Rev. D}
  {\bf 80} (Oct, 2009) 083519.

\bibitem{Lewis:2002ah}
A.~Lewis and S.~Bridle, {\it {Cosmological parameters from CMB and other data:
  A Monte Carlo approach}},  {\em Phys.Rev.} {\bf D66} (2002) 103511,
  [\href{http://xxx.lanl.gov/abs/astro-ph/0205436}{{\tt astro-ph/0205436}}].

\bibitem{Cosmomc}
A.~Lewis and S.~Bridle, {\it {Cosmological MonteCarlo (CosmoMC)}},  2013.
\newblock Publicly available Markov-Chain Monte-Carlo likelihood sampler:
  \texttt{\ifpdf\href{http://cosmologist.info/cosmomc/}{http://cosmologist.info/cosmomc}\else{http://cosmologist.info/cosmomc}\fi}.

\bibitem{Lewis:1999bs}
A.~Lewis, A.~Challinor, and A.~Lasenby, {\it {Efficient Computation of Cosmic
  Microwave Background Anisotropies in Closed Friedmann-Robertson-Walker
  Models}},  {\em Astrophys.J.} {\bf 538} (2000) 473--476,
  [\href{http://xxx.lanl.gov/abs/astro-ph/9911177}{{\tt astro-ph/9911177}}].

\bibitem{Camb}
A.~Lewis and A.~Challinor, {\it {Code for Anisotropies in the Microwave
  Background (CAMB)}},  2013.
\newblock Publicly available CMB-Boltzmann code:
  \texttt{\ifpdf\href{http://www.camb.info/}{http://www.camb.info}\else{http://www.camb.info}\fi}.

\bibitem{Pisanti:2007hk}
O.~Pisanti, A.~Cirillo, S.~Esposito, F.~Iocco, G.~Mangano, et~al., {\it
  {PArthENoPE: Public Algorithm Evaluating the Nucleosynthesis of Primordial
  Elements}},  {\em Comput.Phys.Commun.} {\bf 178} (2008) 956--971,
  [\href{http://xxx.lanl.gov/abs/0705.0290}{{\tt arXiv:0705.0290}}].

\bibitem{Planck:2013kta}
{\bf Planck collaboration} Collaboration, P.~Ade et~al., {\it {Planck 2013
  results. XV. CMB power spectra and likelihood}},
  \href{http://xxx.lanl.gov/abs/1303.5075}{{\tt arXiv:1303.5075}}.

\bibitem{Ade:2013zuv}
{\bf Planck Collaboration} Collaboration, P.~Ade et~al., {\it {Planck 2013
  results. XVI. Cosmological parameters}},
  \href{http://xxx.lanl.gov/abs/1303.5076}{{\tt arXiv:1303.5076}}.

\bibitem{Bennett:2012fp}
{\bf WMAP} Collaboration, C.~Bennett, D.~Larson, J.~Weiland, N.~Jarosik,
  G.~Hinshaw, et~al., {\it {Nine-Year Wilkinson Microwave Anisotropy Probe
  (WMAP) Observations: Final Maps and Results}},  {\em Astrophys.J.Suppl.} {\bf
  208} (2013) 20, [\href{http://xxx.lanl.gov/abs/1212.5225}{{\tt
  arXiv:1212.5225}}].

\bibitem{Story:2012wx}
K.~Story, C.~Reichardt, Z.~Hou, R.~Keisler, K.~Aird, et~al., {\it {A
  Measurement of the Cosmic Microwave Background Damping Tail from the
  2500-square-degree SPT-SZ survey}},  {\em Astrophys.J.} {\bf 779} (2013) 86,
  [\href{http://xxx.lanl.gov/abs/1210.7231}{{\tt arXiv:1210.7231}}].

\bibitem{Das:2013zf}
S.~Das, T.~Louis, M.~R. Nolta, G.~E. Addison, E.~S. Battistelli, et~al., {\it
  {The Atacama Cosmology Telescope: Temperature and Gravitational Lensing Power
  Spectrum Measurements from Three Seasons of Data}},
  \href{http://xxx.lanl.gov/abs/1301.1037}{{\tt arXiv:1301.1037}}.

\bibitem{Percival:2009xn}
{\bf SDSS Collaboration} Collaboration, W.~J. Percival et~al., {\it {Baryon
  Acoustic Oscillations in the Sloan Digital Sky Survey Data Release 7 Galaxy
  Sample}},  {\em Mon.Not.Roy.Astron.Soc.} {\bf 401} (2010) 2148--2168,
  [\href{http://xxx.lanl.gov/abs/0907.1660}{{\tt arXiv:0907.1660}}].

\bibitem{Padmanabhan:2012hf}
N.~Padmanabhan, X.~Xu, D.~J. Eisenstein, R.~Scalzo, A.~J. Cuesta, et~al., {\it
  {A 2
  Methods and Application to the Sloan Digital Sky Survey}},  {\em
  Mon.Not.Roy.Astron.Soc.} {\bf 427} (2012) 2132,
  [\href{http://xxx.lanl.gov/abs/1202.0090}{{\tt arXiv:1202.0090}}].

\bibitem{Beutler:2011hx}
F.~Beutler, C.~Blake, M.~Colless, D.~H. Jones, L.~Staveley-Smith, et~al., {\it
  {The 6dF Galaxy Survey: Baryon Acoustic Oscillations and the Local Hubble
  Constant}},  {\em Mon.Not.Roy.Astron.Soc.} {\bf 416} (2011) 3017--3032,
  [\href{http://xxx.lanl.gov/abs/1106.3366}{{\tt arXiv:1106.3366}}].

\bibitem{Blake:2011en}
C.~Blake, E.~Kazin, F.~Beutler, T.~Davis, D.~Parkinson, et~al., {\it {The
  WiggleZ Dark Energy Survey: mapping the distance-redshift relation with
  baryon acoustic oscillations}},  {\em Mon.Not.Roy.Astron.Soc.} {\bf 418}
  (2011) 1707--1724, [\href{http://xxx.lanl.gov/abs/1108.2635}{{\tt
  arXiv:1108.2635}}].

\bibitem{Anderson:2012sa}
L.~Anderson, E.~Aubourg, S.~Bailey, D.~Bizyaev, M.~Blanton, et~al., {\it {The
  clustering of galaxies in the SDSS-III Baryon Oscillation Spectroscopic
  Survey: Baryon Acoustic Oscillations in the Data Release 9 Spectroscopic
  Galaxy Sample}},  {\em Mon.Not.Roy.Astron.Soc.} {\bf 427} (2013), no.~4
  3435--3467, [\href{http://xxx.lanl.gov/abs/1203.6594}{{\tt
  arXiv:1203.6594}}].

\bibitem{Conley:2011ku}
A.~Conley, J.~Guy, M.~Sullivan, N.~Regnault, P.~Astier, et~al., {\it {Supernova
  Constraints and Systematic Uncertainties from the First 3 Years of the
  Supernova Legacy Survey}},  {\em Astrophys.J.Suppl.} {\bf 192} (2011) 1,
  [\href{http://xxx.lanl.gov/abs/1104.1443}{{\tt arXiv:1104.1443}}].

\bibitem{Riess:2011yx}
A.~G. Riess, L.~Macri, S.~Casertano, H.~Lampeitl, H.~C. Ferguson, et~al., {\it
  {A 3\% Solution: Determination of the Hubble Constant with the Hubble Space
  Telescope and Wide Field Camera 3}},  {\em Astrophys.J.} {\bf 730} (2011)
  119, [\href{http://xxx.lanl.gov/abs/1103.2976}{{\tt arXiv:1103.2976}}].

\bibitem{Kilbinger:2012qz}
M.~Kilbinger, L.~Fu, C.~Heymans, F.~Simpson, J.~Benjamin, et~al., {\it
  {CFHTLenS: Combined probe cosmological model comparison using 2D weak
  gravitational lensing}},  {\em Monthly Notices of the Royal Astronomical
  Society} {\bf 430} (2013), no.~3 2200--2220,
  [\href{http://xxx.lanl.gov/abs/1212.3338}{{\tt arXiv:1212.3338}}].

\bibitem{Ade:2013lmv}
{\bf Planck Collaboration} Collaboration, P.~Ade et~al., {\it {Planck 2013
  results. XX. Cosmology from Sunyaev-Zeldovich cluster counts}},
  \href{http://xxx.lanl.gov/abs/1303.5080}{{\tt arXiv:1303.5080}}.

\bibitem{Bean:2003fb}
R.~Bean and O.~Dore, {\it {Probing dark energy perturbations: The Dark energy
  equation of state and speed of sound as measured by WMAP}},  {\em Phys.Rev.}
  {\bf D69} (2004) 083503,
  [\href{http://xxx.lanl.gov/abs/astro-ph/0307100}{{\tt astro-ph/0307100}}].

\bibitem{dePutter:2010vy}
R.~de~Putter, D.~Huterer, and E.~V. Linder, {\it {Measuring the Speed of Dark:
  Detecting Dark Energy Perturbations}},  {\em Phys.Rev.} {\bf D81} (2010)
  103513, [\href{http://xxx.lanl.gov/abs/1002.1311}{{\tt arXiv:1002.1311}}].

\bibitem{Kunz:2003iz}
M.~Kunz, P.-S. Corasaniti, D.~Parkinson, and E.~J. Copeland, {\it
  {Model-independent dark energy test with sigma(8) using results from the
  Wilkinson microwave anisotropy probe}},  {\em Phys.Rev.} {\bf D70} (2004)
  041301, [\href{http://xxx.lanl.gov/abs/astro-ph/0307346}{{\tt
  astro-ph/0307346}}].

\bibitem{Hollenstein:2009ph}
L.~Hollenstein, D.~Sapone, R.~Crittenden, and B.~M. Schaefer, {\it {Constraints
  on early dark energy from CMB lensing and weak lensing tomography}},  {\em
  JCAP} {\bf 0904} (2009) 012, [\href{http://xxx.lanl.gov/abs/0902.1494}{{\tt
  arXiv:0902.1494}}].

\bibitem{Amendola:2004wa}
L.~Amendola and C.~Quercellini, {\it {Skewness as a test of the equivalence
  principle}},  {\em Phys.Rev.Lett.} {\bf 92} (2004) 181102,
  [\href{http://xxx.lanl.gov/abs/astro-ph/0403019}{{\tt astro-ph/0403019}}].

\bibitem{Linder:2005in}
E.~V. Linder, {\it {Cosmic growth history and expansion history}},  {\em
  Phys.Rev.} {\bf D72} (2005) 043529,
  [\href{http://xxx.lanl.gov/abs/astro-ph/0507263}{{\tt astro-ph/0507263}}].

\bibitem{Pogosian:2010tj}
L.~Pogosian, A.~Silvestri, K.~Koyama, and G.-B. Zhao, {\it {How to optimally
  parametrize deviations from General Relativity in the evolution of
  cosmological perturbations?}},  {\em Phys.Rev.} {\bf D81} (2010) 104023,
  [\href{http://xxx.lanl.gov/abs/1002.2382}{{\tt arXiv:1002.2382}}].

\bibitem{Bean:2010zq}
R.~Bean and M.~Tangmatitham, {\it {Current constraints on the cosmic growth
  history}},  {\em Phys.Rev.} {\bf D81} (2010) 083534,
  [\href{http://xxx.lanl.gov/abs/1002.4197}{{\tt arXiv:1002.4197}}].

\end{thebibliography}\endgroup
\bibliographystyle{JHEP}

\end{document}